\documentclass[traditabstract]{aa}

\usepackage{txfonts}
\usepackage{graphicx}
\usepackage{amssymb}

\newcommand\T{\rule{0pt}{2.0ex}}
\newcommand\B{\rule[-0.8ex]{0pt}{0pt}}
\newcommand\BB{\rule[-1.0ex]{0pt}{0pt}}

\usepackage{natbib}
\bibpunct{(}{)}{;}{a}{}{,}

\title{Limb darkening laws for two exoplanet host stars derived from 3D stellar model atmospheres}
\subtitle{Comparison with 1D models and HST light curve observations}

\author{
W. Hayek\inst{1,3}
\and D. Sing\inst{1}
\and F. Pont\inst{1}
\and M. Asplund\inst{2,3}
}
\institute{
Astrophysics Group, School of Physics, University of Exeter, Stocker Road, Exeter, EX4 4QL\\
\email{[hayek,sing,fpont]@astro.ex.ac.uk}
\and
Research School of Astronomy \& Astrophysics, Cotter Road, Weston Creek 2611, Australia\\
\email{martin@mso.anu.edu.au}
\and
Max Planck Institut f\"ur Astrophysik, Karl-Schwarzschild-Str. 1, D-85741 Garching, Germany\\
}

\date{\today}

\abstract{
We compare limb darkening laws derived from 3D hydrodynamical model atmospheres and 1D hydrostatic \texttt{MARCS} models for the host stars of two well-studied transiting exoplanet systems, the late-type dwarfs \object{HD~209458} and \object{HD~189733}. The surface brightness distribution of the stellar disks is calculated for a wide spectral range using 3D LTE spectrum formation and opacity sampling\thanks{Full theoretical spectra for both stars are available in electronic form at the CDS via anonymous ftp to cdsarc.u-strasbg.fr (130.79.128.5) or via http://cdsweb.u-strasbg.fr/cgi-bin/qcat?J/A+A/, as well as at www.astro.ex.ac.uk/people/sing.}. We test our theoretical predictions using least-squares fits of model light curves to wavelength-integrated primary eclipses that were observed with the Hubble Space Telescope (HST).

The limb darkening law derived from the 3D model of HD~209458 in the spectral region between $2900$\,{\AA} and $5700$\,{\AA} produces significantly better fits to the HST data, removing systematic residuals that were previously observed for model light curves based on 1D limb darkening predictions. This difference arises mainly from the shallower mean temperature structure of the 3D model, which is a consequence of the explicit simulation of stellar surface granulation where 1D models need to rely on simplified recipes. In the case of HD~189733, the model atmospheres produce practically equivalent limb darkening curves between $2900$\,{\AA} and $5700$\,{\AA}, partly due to obstruction by spectral lines, and the data are not sufficient to distinguish between the light curves. We also analyze HST observations between $5350$\,{\AA} and $10500$\,{\AA} for this star; the 3D model leads to a better fit compared to 1D limb darkening predictions.

The significant improvement of fit quality for the HD~209458 system demonstrates the higher degree of realism of 3D hydrodynamical models and the importance of surface granulation for the formation of the atmospheric radiation field of late-type stars. This result agrees well with recent investigations of limb darkening in the solar continuum and other observational tests of the 3D models. The case of HD~189733 is no contradiction as the model light curves are less sensitive to the temperature stratification of the stellar atmosphere and the observed data in the $2900$\,{\AA} - $5700$\,{\AA} region are not sufficient to distinguish more clearly between the 3D and 1D limb darkening predictions.
}


\begin{document}
\maketitle

\section{Introduction}

The surface brightness of a star varies across its visible disc as the optical path length along the line of sight depends on the viewing angle onto the atmosphere. Closer to the limb, radiative flux emerges from higher and cooler regions in the stellar photosphere, therefore causing ``limb darkening''. This angular dependence is commonly described with an analytical, linear or nonlinear law, which forms an essential ingredient for analyzing the light curves of transiting exoplanets and eclipsing binary stars. The occulting body indirectly samples radiative intensities in the observer's direction as it travels across the stellar disk, leaving a clear signature in the shape of the light curve. A uniform intensity distribution would result in an almost rectangular shape with a flat-bottom light curve; the limb darkening effect causes shallower slopes during the ingress and egress phases and a curved signal in the center of the transit.

Light curve analyses rely on theoretical predictions for limb darkening in general. With few exceptions, these are still mostly derived from classical 1D hydrostatic stellar model atmospheres, for which extensive grids are available \citep[see, e.g.,][]{Claret:2000,Sing:2010}. Very high signal-to-noise light curve observations were obtained with the Hubble Space Telescope (HST) for HD~209458 \citep{Brownetal:2001,Knutsonetal:2007} and for HD~189733 \citep{Pontetal:2008,Singetal:2011}. The data quality enables a direct fit of at least a low-order limb darkening law \citep[see, e.g., the discussion in][]{Southworth:2008} along with the other transit parameters, and therefore an assessment of the accuracy of model predictions for limb darkening. It turns out that using 1D models leads to substantially higher residuals in the light curve fits compared to directly fitted limb darkening laws. Moreover, these residuals exhibit deviations in the ingress and egress phases of the transit that are characteristic for incorrect limb darkening \citep{Knutsonetal:2007}. Accurate independent model predictions for limb darkening are nevertheless generally preferable to fitting a law to the data in order to reduce the number of free parameters in the fit and avoid the lower accuracy of the linear or quadratic laws that are typically used in the method. This would also eliminate the wavelength-dependent degeneracy of limb darkening with transit depth and thus with the important planet-to-star radius ratio on which transit spectroscopy of planetary atmospheres is based.

As the limb darkening curve is mainly determined by the photospheric temperature gradient, the systematic residuals in the light curve fits of \citet{Knutsonetal:2007} for HD~209458b point towards shortcomings in the structure of 1D stellar model atmospheres. \citet{Claret:2009} further investigated the case, comparing limb darkening coefficients derived from various 1D model atmospheres and different analytical laws with coefficients that resulted from empirical fits by \citet{Southworth:2008}, and concluded that current 1D atmosphere models of HD~209458 cannot produce consistent results.

The Sun has traditionally been the most important bench mark for testing the realism of model atmospheres as, e.g., the surface brightness distribution can be readily measured in great detail \citep{Neckeletal:1994}. Comparisons with different solar model atmospheres by \citet{Asplundetal:2009} revealed that the latest generation of 3D hydrodynamical models is capable of satisfying this important test with very good accuracy (see their Fig.~2), while the limb darkening predicted by 1D hydrostatic \texttt{MARCS} models is too strong; \citet{Singetal:2008} found a similar result for a solar 1D \texttt{ATLAS} model. 3D models take the effect of convective motions in the surface granulation explicitly into account and are thus able to reproduce the solar atmosphere with a higher degree of realism; see \citet{Nordlundetal:2009} for a detailed description of the physics and methods.

We compute detailed theoretical surface intensity distributions from 3D hydrodynamical model atmospheres of HD~209458 and HD~189733 across a large wavelength range and derive limb darkening coefficients for the spectral band between $2900$\,{\AA} and $5700$\,{\AA} covered by the HST STIS G430L grating, where limb darkening is more sensitive to the photospheric temperature gradient, using least-squares fits of a nonlinear law to the numerical calculations. We additionally investigate the region between $5350$\,{\AA} and $10500$\,{\AA} using ACS HRC G800L data in the case of HD~189733, as our STIS observations include only a small fraction of the transit. The models and the choice of stellar parameters are described in Sect.~\ref{sec:atmomodels}, followed by a discussion of the 3D radiative transfer computations needed for obtaining theoretical spectra in Sect.~\ref{sec:specld}. We compare our predictions of 3D limb darkening with the results from 1D \texttt{MARCS} models in Sect.~\ref{sec:3D1Dlimbdark} and test their accuracy using fits to HST transiting light curves in Sect.~\ref{sec:lcfits}. The paper concludes with a summary of our results in Sect.~\ref{sec:conclusions}. Limb darkening coefficients for a selection of various standard broadband filters and instrument sensitivities are presented in Appendix~\ref{sec:ldccompilation}.

\section{Stellar model atmospheres}\label{sec:atmomodels}

\begin{figure*}[htbp]
\centering
\includegraphics[width=8.5cm]{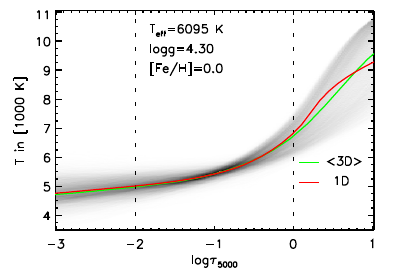}
\includegraphics[width=8.5cm]{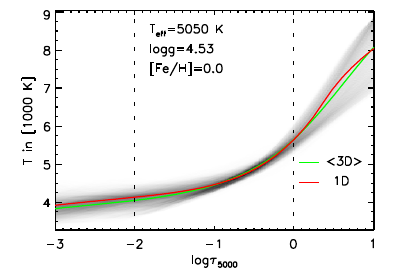}
\caption{Temperature distributions (gray shades) on surfaces with constant optical depth at $5000$\,{\AA} for arbitrary snapshots of the 3D models for HD~209458 (left) and HD~189733 (right). The spatial and temporal mean temperature stratification of the 3D models (green solid lines) is compared to 1D \texttt{MARCS} models (red solid lines) with the same stellar parameters. The dashed lines bracket the approximate height region between $\log\tau_{5000}=0$ and $\log\tau_{5000}=-2$, from which the dominant contribution to the continuum surface brightness between the disk center at $\mu=1.0$ and the limb near $\mu=0.01$ is emitted.}
\label{fig:3Dmodel}
\end{figure*}

Theoretical models of stellar atmospheres have traditionally been based on the approximation of horizontal homogeneity and plane parallel or spherically symmetric geometry. The problem is thereby reduced to one dimension, which enables very fast computation on modern computers and a very detailed treatment of radiative energy transfer. Large grids of stellar atmospheres were created, such as the Kurucz grid \citep{Kurucz:1996}, the MARCS grid \citep{Gustafssonetal:2008} and the NextGen grid \citep{Hauschildtetal:1999}.

However, the atmospheres of late-type stars are known to develop a convection zone in the envelope that reaches the stellar surface, producing a granulation pattern that causes strong horizontal inhomogeneities in the atmospheric structure, which strongly influences the outgoing radiation field \citep[see the discussion in, e.g.,][for the important case of the Sun]{Nordlundetal:2009}. 3D time-dependent radiation-hydrodynamical model atmospheres take these convective motions explicitly into account.

We construct two 3D models to represent the exoplanet host stars HD~209458 and HD~189733 using the \texttt{Stagger Code} \citep{Nordlundetal:1995}. The coupled equations of compressible hydrodynamics and time-independent radiative transfer are solved simultaneously in a plane-parallel box with a resolution of $240\times240\times240$ grid points, covering $\sim10$ granules at a time to represent stellar surface convection. The dependence of continuous and spectral line opacities on wavelength and atmospheric height is approximated using the opacity binning method \citep{Nordlund:1982}. Opacities were taken from the same opacity package that was used for the spectrum calculations (see Sect.~\ref{sec:specld}) and sorted into 12 groups as a function of wavelength and Rosseland optical depth of their monochromatic optical surfaces. Based on the spatial and temporal mean stratification of the 3D model, intensity-mean and Rosseland mean opacities are computed for each group. They are subsequently combined into a single opacity using exponential bridging as a function of optical depth; see \citet{Skartlien:2000} for a detailed description of the method and \citet{Hayeketal:2010} for a summary of opacity sources. The chemical composition is based on the solar abundances of \citet{Asplundetal:2005}. The simulations reach a horizontal extent of $12\,\mathrm{Mm}\times12\,\mathrm{Mm}$ (HD~209458) and $4\,\mathrm{Mm}\times4\,\mathrm{Mm}$ (HD~189733) on equidistant axes with periodic boundaries. On the vertical axis, which has finer resolution around the continuum optical surface of the models, $5.9$\,Mm (HD~209458) and $2.2$\,Mm (HD~189733) are covered with open boundaries to allow free inflow and outflow of stellar gas. This corresponds to an optical depth range of about $10^{-6}\lesssim\tau_{5000}\lesssim10^{7}$ at $5000$\,{\AA} for both models, thus including all atmospheric layers that are relevant for spectrum formation above the optical surface and for surface granulation below. The simulations assume magnetically quiet convection, which ignores the effects of magnetic fields on plasma dynamics. This simplification is valid for stars like the Sun or HD~209458, where this activity should only weakly influence the outgoing radiation field in most parts of the spectrum. The case of HD~189733, where transit light curves show the existence of numerous star spots \citep[see][]{Singetal:2011}, may require more scrutiny in the future with larger models that allow the inclusion of star spots.

The 1D atmosphere models are based on interpolations in the \texttt{MARCS} grid; see \citet{Gustafssonetal:2008} for a description of the physics, input data and methods.

In order to test the consistency of stellar parameters, we perform LTE (local thermodynamic equilibrium) abundance analyses with lines of neutral and ionized iron on high resolution, high signal-to-noise HARPS spectra\footnote{Based on data obtained from the ESO Science Archive Facility.}. The observed equivalent widths, derived from least-squares fits of Voigt profiles to the data, are compared to synthetic LTE line profiles that we compute with the \texttt{SCATE} line formation code \citep[for a description see][]{Hayeketal:2011}. \texttt{SCATE} uses the same opacity package on which the \texttt{MARCS} grid and the 3D model atmospheres are based. The code was also used to compute the intensity spectra needed for deriving the limb darkening laws (see Sect.~\ref{sec:specld}), ensuring complete consistency in our investigation.

We base our abundance analysis on the \ion{Fe}{I} line list of \citet{Asplundetal:2000}, with line data taken from \citet{Blackwelletal:1995} and \citet{Holwegeretal:1995}, as well as on the compilation of \ion{Fe}{II} laboratory data of \citet{Melendezetal:2009}. The effective temperature is constrained by removing trends of the \ion{Fe}{I} abundance with excitation potential $\chi_{\mathrm{ex}}$, where low excitation lines exhibit the strongest sensitivity to temperature due to the excitation-ionization balance. The correct surface gravity is found by requiring matching abundances of \ion{Fe}{I} and \ion{Fe}{II} through LTE ionization equilibrium, where \ion{Fe}{II} lines exhibit the strongest sensitivity. It is well known that departures from LTE affect absorber populations in stellar atmospheres, in particular in the case of neutral iron. These effects should be small for ionized iron, which thus yields a more reliable abundance measurement \citep[see the review of][]{Asplund:2005}. For our goal of deriving theoretical limb darkening coefficients from LTE spectrum formation, achieving LTE ionization equilibrium for \ion{Fe}{I} and \ion{Fe}{II} still leads to consistent results: the observed spectra of HD~209458 and HD~189733 constrain the atmospheric stratification of the models through their stellar parameters, while the same radiative transfer methods and opacity data are then used in turn to predict the stellar surface brightness distribution that enters the model light curves.

\begin{table}
\caption{Abundances of neutral and ionized iron lines, derived from 3D hydrodynamical and 1D hydrostatic model atmospheres.}
\begin{center}
\begin{tabular}{lrrr}
\hline
\hline
\multicolumn{4}{c}{HD~209458 \T}\\
\hline
Data set & $N_{\mathrm{lines}}$ & $\log\epsilon(\mathrm{Fe})^{\mathrm{3D}}$ \T & $\log\epsilon(\mathrm{Fe})^{\mathrm{1D}}$\\
\hline
\T\ion{Fe}{I} B$^{1}$ & 15 		& $7.45\pm0.06$	& $7.40\pm0.05$\\
\ion{Fe}{I} H$^{2}$ & 23 		& $7.43\pm0.07$	& $7.37\pm0.06$\\
\ion{Fe}{II}$^{3}$ & 16 	& $7.50\pm0.10$	& $7.38\pm0.07$\\
\hline\\
\hline
\hline
\multicolumn{4}{c}{HD~189733 \T}\\
\hline
Data set & $N_{\mathrm{lines}}$ & $\log\epsilon(\mathrm{Fe})^{\mathrm{3D}}$ \T & $\log\epsilon(\mathrm{Fe})^{\mathrm{1D}}$\\
\hline
\T\ion{Fe}{I} B$^{1}$ & 11 	& $7.56\pm0.06$	& $7.49\pm0.07$\\
\ion{Fe}{I} H$^{2}$ & 16 	& $7.49\pm0.08$	& $7.43\pm0.08$\\
\ion{Fe}{II}$^{3}$ & 4 	& $7.50\pm0.06$	& $7.46\pm0.06$\\
\hline
\end{tabular}
\tablebib{
(1) \citet{Blackwelletal:1995}; (2) \citet{Holwegeretal:1995}; (3)~\citet{Melendezetal:2009}
}
\end{center}
\label{tab:feabund}
\end{table}

\begin{table*}[htdp]
\caption{Literature values compared to our adopted effective temperature, surface gravity, and metallicity.}
\begin{center}
\begin{tabular}{cccll}
\hline
\hline
\multicolumn{5}{c}{HD~209458 \T}\\
\hline
\T\B $T_{\mathrm{eff}}$ [K] & $\log g$ [cm\,s$^{-2}$] & [Fe/H] & Reference & Method \\
\hline
$6000$	& $4.25\phantom{0}$	& $\phantom{-}0.00$ 	&  \citet{Mazehetal:2000}				& spectroscopic\\
$6063$	& $4.38\phantom{0}$	& $\phantom{-}0.04$		& \citet{Gonzalezetal:2001}			& spectroscopic\\
$6080$	& $4.33\phantom{0}$	& $-0.06$ 				& \citet{Mashonkinaetal:2001}	& spectroscopic\\
$5987$	& $4.24\phantom{0}$	& $-0.04$				& \citet{Sadakaneetal:2002}			& spectroscopic\\
$6100$	& $4.50\phantom{0}$	& $-0.02$				& \citet{Heiteretal:2003}				& spectroscopic\\
$6117$	& $4.48\phantom{0}$	& $\phantom{-}0.02$		& \citet{Santosetal:2004}				& spectroscopic\\
$5993$	&					& 					& \citet{Ramirezetal:2004}			& photometric\\
$6099$	& $4.38\phantom{0}$	& $\phantom{-}0.02$		& \citet{Valentietal:2005}				& spectroscopic\\
$6118$	& $4.50\phantom{0}$	& $\phantom{-}0.03$		& \citet{Sousaetal:2008}				& spectroscopic\\
		& $4.368$				&					& \citet{Southworth:2009}				& light curve\\
$6108$	& $4.33\phantom{0}$	& $-0.04$				& \citet{Casagrandeetal:2011}			& photometric\\
\hline
$6095$	& $4.3\phantom{00}$ 	& $\phantom{-}0.00$		& \T\B adopted (3D model)	& spectroscopic\\
$6000$	& $4.3\phantom{00}$ 	& $\phantom{-}0.00$		& adopted (1D model)	& spectroscopic\\
\hline\\
\hline
\hline
\multicolumn{5}{c}{HD~189733 \T}\\
\hline
\T\B $T_{\mathrm{eff}}$ [K] & $\log g$ [cm\,s$^{-2}$] & [Fe/H] & Reference & Method \\
\hline
$5050$	& $4.53\phantom{0}$	& $-0.03$		& \citet{Bouchyetal:2005}		& spectroscopic\\
$4939$	&  					&			& \citet{vanBelleetal:2009}		& photometric\\
		& $4.610$				&			& \citet{Southworth:2010}		& light curve\\
$5022$	& $4.58\phantom{0}$	& $-0.13$		& \citet{Casagrandeetal:2011}	& photometric\\
\hline
$5050$	& $4.53\phantom{0}$	& $\phantom{-}0.00$ 	& \T\B adopted (3D model)		& spectroscopic\\
$5000$	& $4.53\phantom{0}$	& $\phantom{-}0.00$ 	& adopted (1D model)		& spectroscopic\\
\hline
\end{tabular}
\tablefoot{
Quoted metallicities could not be corrected for differences in the assumed solar composition, as the latter was not always specified in the cited publications, leading to artificial offsets in [Fe/H].
}
\end{center}
\label{tab:stellp}
\end{table*}

A large number of parameter determinations for the G-dwarf HD~209458 are available in the literature (see upper panel in Table~\ref{tab:stellp}). Most of the effective temperature measurements are based on spectroscopic methods similar to our determination. Inherent inaccuracies of the method (uncertainties typically reach the order of $\pm150$\,K) produce some scatter with values ranging between approximately $5990$\,K and $6120$\,K. We obtain a flat distribution of \ion{Fe}{I} abundances with increasing excitation level for a 3D model with time-averaged $\left<T_{\mathrm{eff}}\right>=6095$\,K, which agrees well with the majority of spectroscopic results and with the recent photometric measurement of \citet{Casagrandeetal:2011}. The 1D model yields a smaller effective temperature of $T_{\mathrm{eff}}=6000$\,K. This deviation between the 3D and 1D \texttt{MARCS} results stems from their different temperature gradients and the inhomogeneities of the 3D model that arise from the convective motions. The left panel in Fig.~\ref{fig:3Dmodel} compares the spatial and temporal average temperature stratification of the 3D model as a function of optical depth at $5000$\,{\AA} (green line) with the corresponding $T$-$\tau$ relation of a 1D \texttt{MARCS} model (red line) with the same parameters as the 3D model, which exhibits a steeper temperature gradient around the optical surface at $\log\tau_{5000}=0$. The gray shades in the plot show the temperature distribution of an arbitrary snapshot of the 3D simulation. The dashed lines mark the approximate region that is relevant for the surface brightness distribution across the stellar disk in the continuum (see Sect.~\ref{sec:3D1Dlimbdark}).

Choosing a surface gravity of $\log g=4.3$ leads to reasonably consistent ionization equilibria for both the 3D and 1D models (upper panel of Table~\ref{tab:feabund}) and is in good agreement with the majority of literature results, considering a measurement uncertainty of at least $\pm0.1$ in $\log g$. \cite{Southworth:2009} found $\log g=4.368$ from a light curve analysis, while \citet{Casagrandeetal:2011} derived $\log g=4.33$ from a \emph{Hipparcos} parallax. \citet{Asplundetal:2005} recommend a solar iron abundance of $\log\epsilon(\mathrm{\ion{Fe}{II}})_{\sun}=7.45$, which is sufficiently close to our 3D and 1D \ion{Fe}{II} results for HD~209458, given the uncertainties in $\log g$ and the rather small number of \ion{Fe}{II} lines; we therefore adopt their solar composition for the model atmospheres and spectrum computations.

We obtain a microturbulence parameter $\xi=1.4$\,km\,s$^{-1}$ for the 1D case from the distribution of line-to-line abundances as a function of equivalent width; a \texttt{MARCS} model atmosphere with slightly lower microturbulence ($\xi=1.0$\,km\,s$^{-1}$) is used for all 1D spectrum computations to ensure consistency with the opacity sampling tables (see the discussion in Sect.~\ref{sec:specld}). 3D line formation calculations do not require microturbulence as its contribution to profile broadening is naturally produced by 3D model atmospheres through Doppler shifts in the convective plasma flow.

To our knowledge, three independent spectroscopic and photometric measurements of the effective temperature of the K-dwarf HD~189733 exist in the literature (see lower panel in Table~\ref{tab:stellp}), which deviate by up to about $110$\,K and are therefore in reasonable agreement considering the systematic effects of the different methods. We determine $\left<T_{\mathrm{eff}}\right>=5050$\,K for our 3D model, which is coincidentally equal to the spectroscopic value of \citet{Bouchyetal:2005} that was derived with a 1D \texttt{ATLAS} model. The 1D \texttt{MARCS} model produces a slightly lower effective temperature of $T_{\mathrm{eff}}=5000$\,K.

We find good agreement for our 3D ionization equilibrium (see lower panel of Table~\ref{tab:feabund}) for the surface gravity $\log g=4.53$ of \citet{Bouchyetal:2005}, which is also compatible with the light curve fit of \citet{Southworth:2010} and the parallax-based result of \citet{Casagrandeetal:2011}. Similar agreement is found for the 1D model with its effective temperature of $5000$\,K. The accuracy of our $\log g$ determinations is limited due to the very low number of usable \ion{Fe}{II} lines that we measured in the spectrum, but the stellar surface gravity is only of secondary importance for the atmospheric temperature gradient of the model. The solar iron abundance of \citet{Asplundetal:2005} agrees again reasonably well with our measurement. We determine a microturbulence parameter of $\xi=0.9$\,km\,s$^{-1}$; we therefore choose again a grid model with $\xi=1.0$\,km\,s$^{-1}$ for the 1D calculations.

\section{LTE spectrum formation and limb darkening laws}\label{sec:specld}

\begin{figure*}[htbp]
\centering
\includegraphics[width=8.5cm]{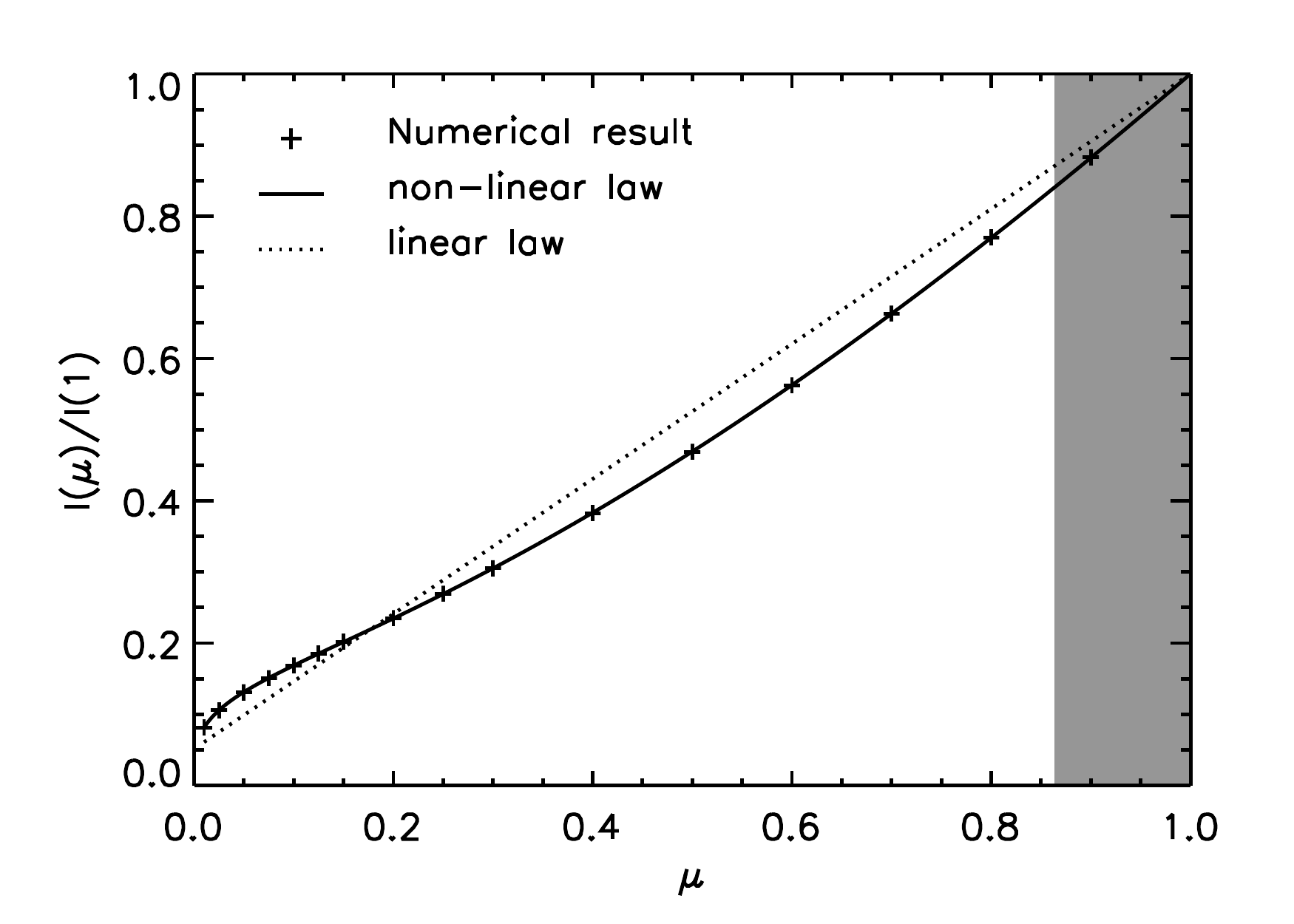}
\includegraphics[width=8.5cm]{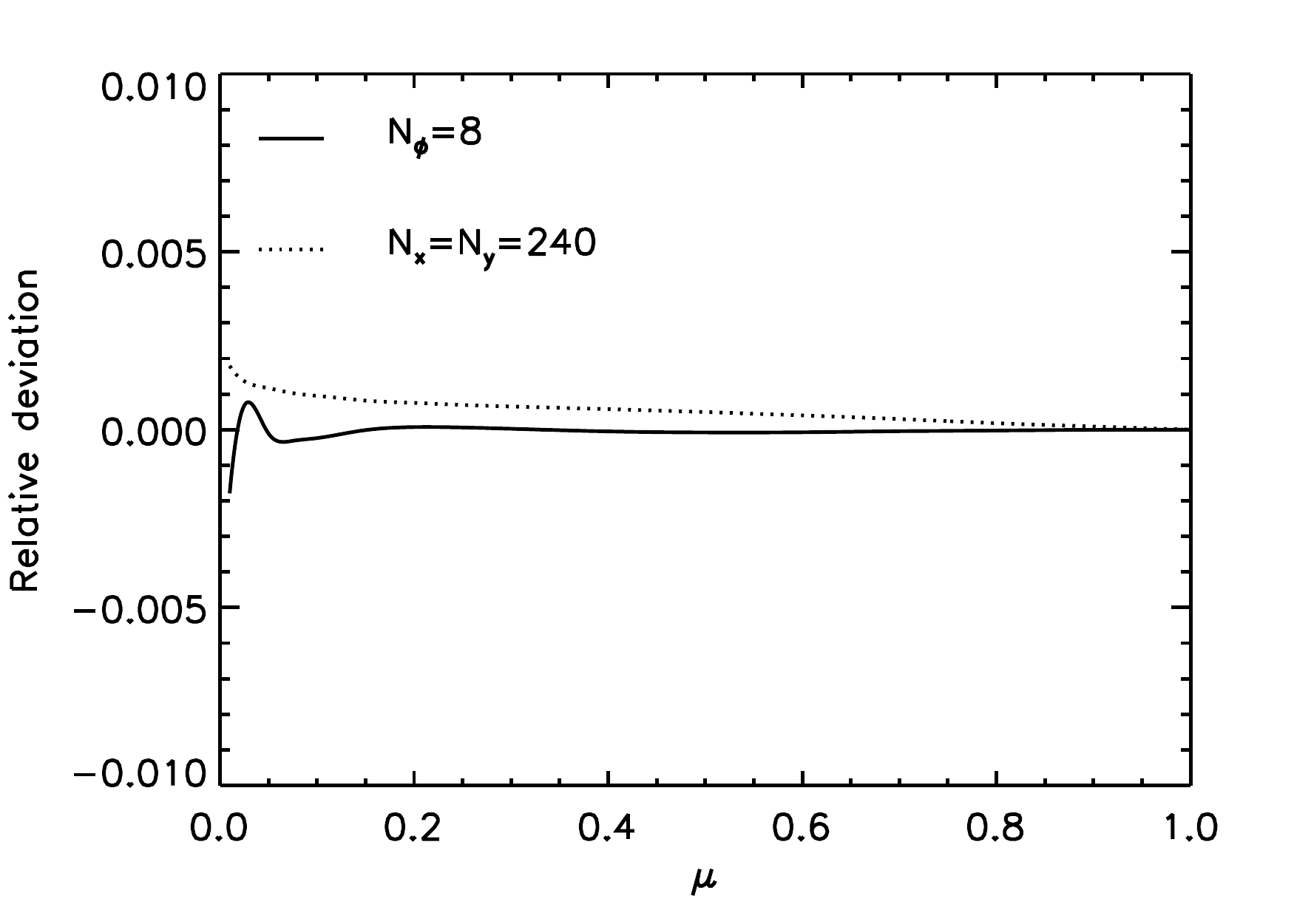}
\caption{\textit{Left:} Theoretical average surface brightness distribution as a function of projection factor $\mu$ integrated over the spectrum between $3400$\,{\AA} and $3500$\,{\AA}, computed using our ``standard'' numerical resolution for an arbitrary snapshot of the 3D model of HD~189733 (crosses), and least-squares fits of the \citet{Claret:2000} nonlinear law (solid line) and a linear law (dotted line). The gray-shaded area indicates the central part of the stellar disk that is not reached during the transit. \textit{Right:} Relative deviation of spline interpolations of theoretical limb darkening computed with $N_{\phi}=8$ direction angles (solid line) and with the full horizontal resolution of $N_{x}=N_{y}=240$ grid points (dotted line) from the interpolated ``standard'' setting.}
\label{fig:ldtestldphirescompare}
\end{figure*}

\begin{figure*}[htdp]
\centering
\includegraphics[width=8.5cm]{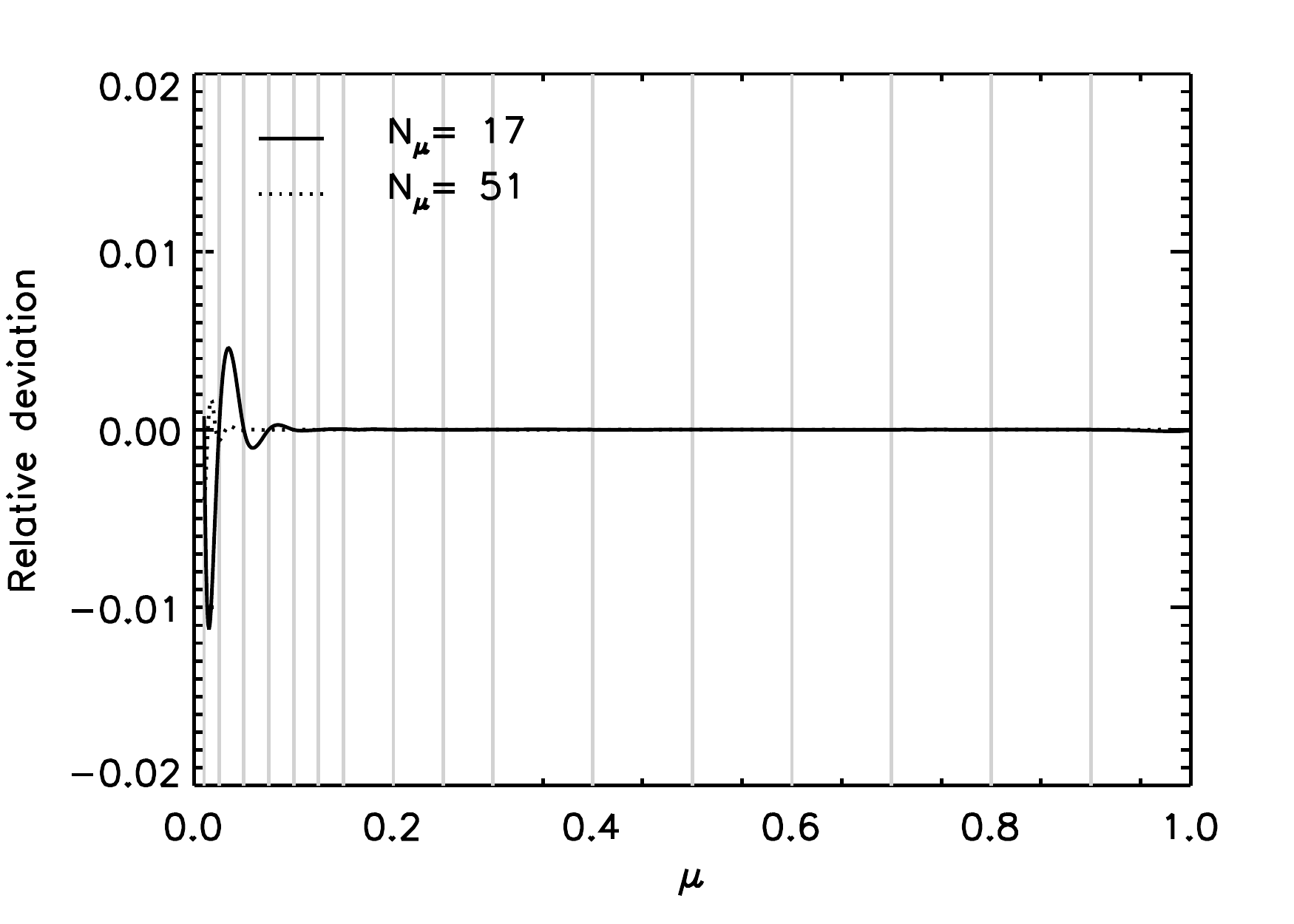}
\includegraphics[width=8.5cm]{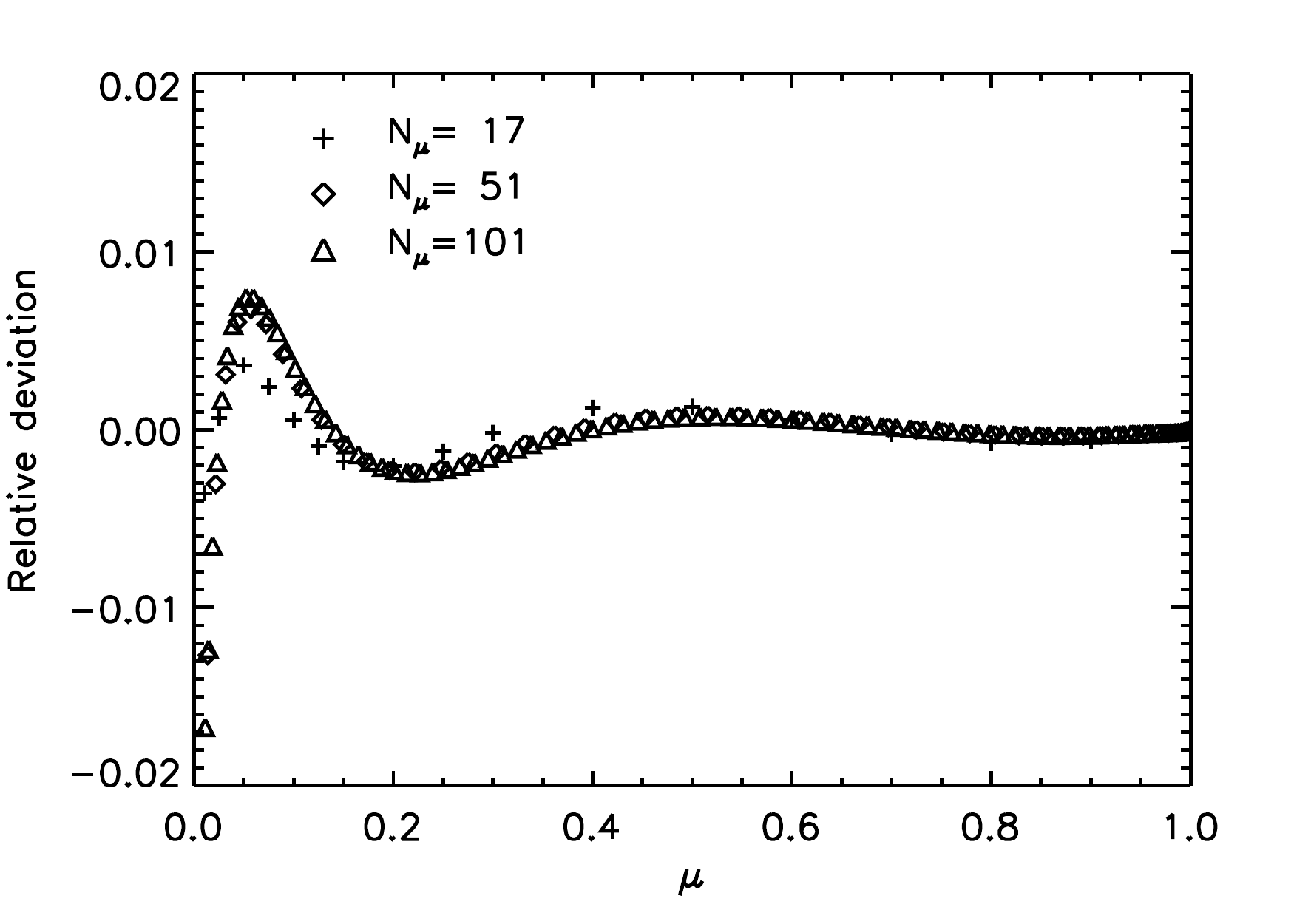}
\caption{\textit{Left:} Relative deviation of the spline-interpolated limb darkening predictions for $N_{\mu}=17$ (solid line) and $N_{\mu}=51$ (dotted line) from the case with $N_{\mu}=101$ as a function of $\mu$. Grey vertical lines mark the position of the 17 $\mu$ angles used in the ``standard'' resolution. \textit{Right:} Relative deviation of the least-squares-fitted limb darkening laws from the computed data points for $N_{\mu}=17$ (crosses), $N_{\mu}=51$ (diamonds) and $N_{\mu}=101$ (triangles) as a function of $\mu$. All calculations are based on the 3D model of HD~189733.}
\label{fig:ldmucompare}
\end{figure*}

\begin{figure*}[htdp]
\centering
\includegraphics[width=8.5cm]{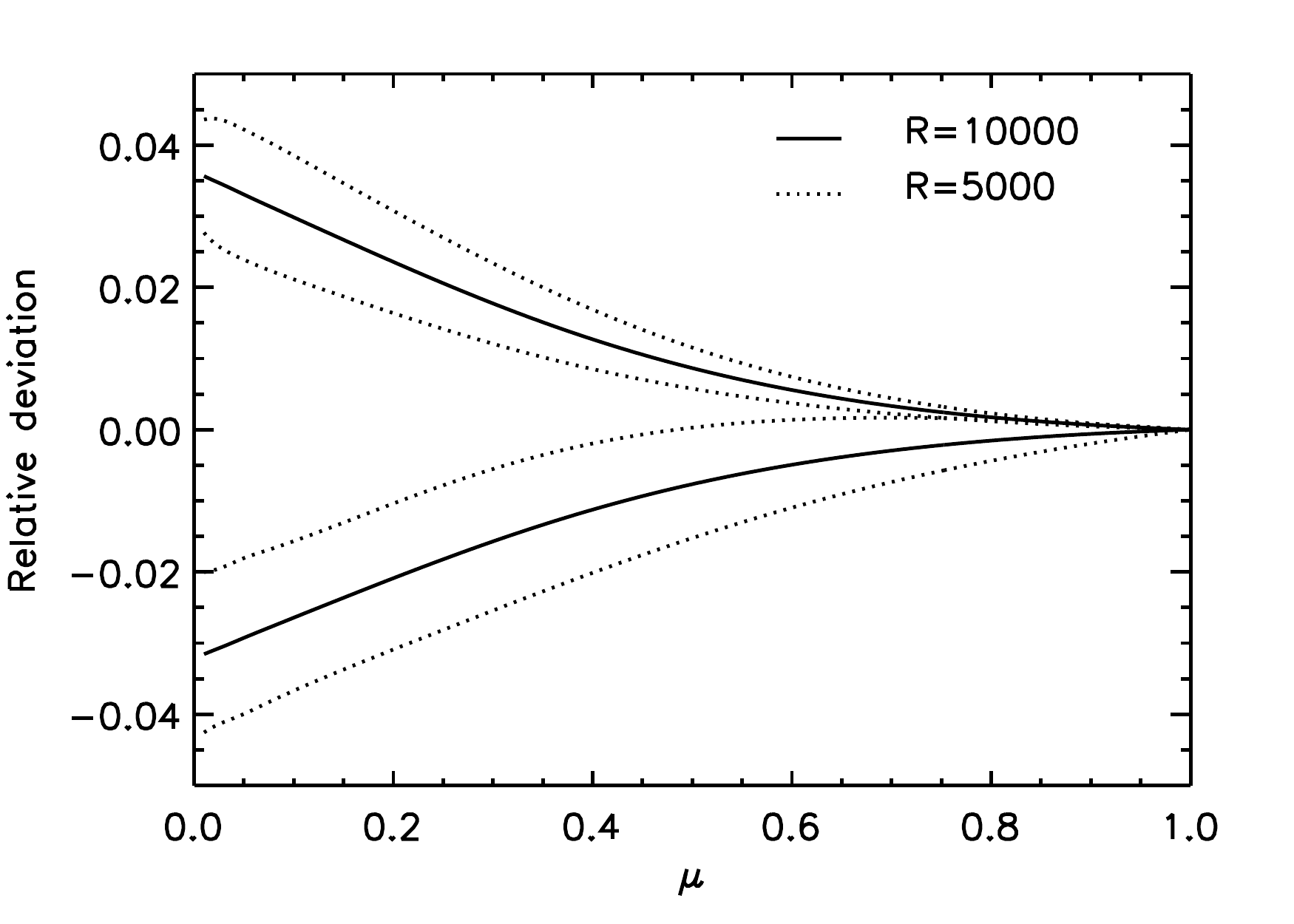}
\includegraphics[width=8.5cm]{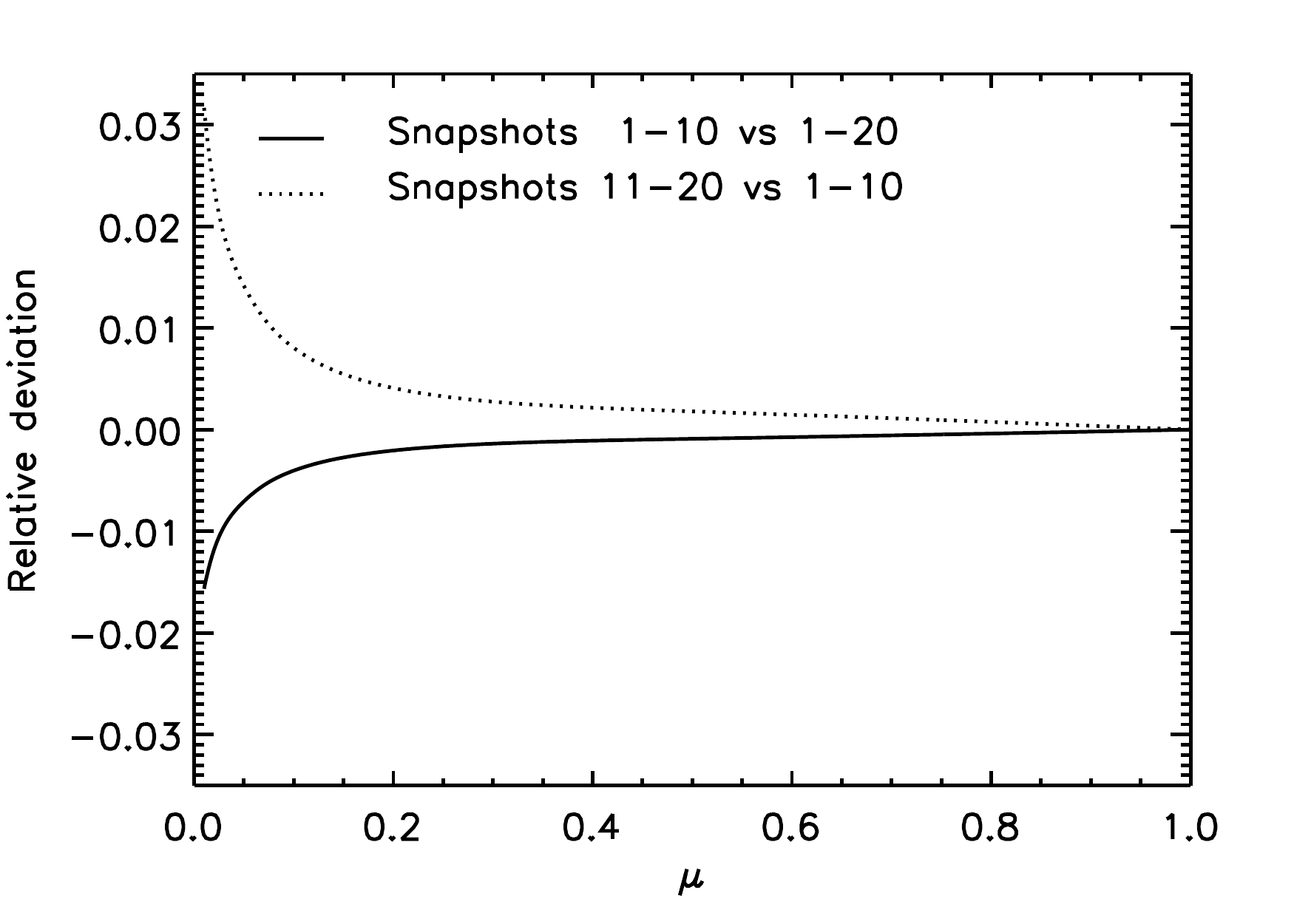}
\caption{\textit{Left:} Relative deviation of the spline-interpolated limb darkening predictions computed for a single snapshot of the 3D model for HD~189733 with reduced wavelength sampling of $R=10000$ (solid lines) and $R=5000$ (dotted lines) from the full sampling with $R=20000$. Each set of resampled wavelengths is shown in either case, resulting in two or four choices, respectively. \textit{Right:} Relative deviation of the spline-interpolated limb darkening predictions computed for the first 10 snapshots of a time sequence with $N_{t}=20$ (solid line) and deviation of the predictions computed for the first and last 10 snapshots of the same sequence (dotted line). All calculations are based on the 3D model of HD~189733.}
\label{fig:ldossequencecompare}
\end{figure*}

\begin{table*}[htdp]
\caption{Comparison of 4th order limb darkening coefficients, goodness of fit, radiative fluxes from the nonlinear law and numerical quadrature, and their relative deviation for various choices of $N_{\mu}$.}
\begin{center}
\begin{tabular}{ccccccccc}
\hline\hline
$N_{\mu}$ & $c_{1}$ & $c_{2}$ & $c_{3}$ & $c_{4}$ & $\chi^{2}_{\nu}$ \BB\T & $F_{\mathrm{nl}}/\pi I(1)$ & $F_{\mathrm{num}}/\pi I(1)$ & $\left|\Delta F_{\mathrm{rel}}\right|$\\
\hline
\phantom{1}17	&  0.5999 & $-1.0136$ &  1.7021 & $-0.3179$ & $1.95\cdot10^{-7}$ & 0.64738 & 0.65040 & $4.64\cdot10^{-3}$ \T \\
\phantom{1}51	&  0.6479 & $-1.1686$ &  1.8951 & $-0.3994$ & $3.51\cdot10^{-7}$ & 0.64747 & 0.64751 & $5.97\cdot10^{-5}$ \\
101	&  0.6548 & $-1.1911$ &  1.9230 & $-0.4111$ & $4.34\cdot10^{-7}$ & 0.64748 & 0.64751 & $4.54\cdot10^{-5}$ \\
\hline
\end{tabular}
\tablefoot{
All calculations are based on the 3D model of HD~189733 and integrated over the wavelength band between $3400$\,{\AA} and $3500$\,{\AA}. Radiative fluxes are divided by $\pi I(1)$ to obtain the same normalization as the limb darkening laws.
}
\end{center}
\label{tab:ldtest}
\end{table*}

We use the \texttt{SCATE} line formation code to compute surface intensities for all models at different angles with respect to the vertical axis, corresponding to varying positions on the stellar disk. The calculations are based on continuous opacities from \citet{Gustafssonetal:1975} with updates by R. Trampedach (priv. comm.), as well as sampled line opacities from B. Plez (priv. comm.); see also \citet{Gustafssonetal:2008}. The chemical composition assumes the solar abundances of \citet{Asplundetal:2005}. In order to reduce the computational expense of full spectrum calculations in 3D geometry, all opacities were precomputed and tabulated as a function of wavelength, pressure and temperature using the approximation of local thermal equilibrium (LTE) and neglecting the effects of scattering. This approach precludes a detailed treatment of Doppler shifts, contrary to our 3D calculations of iron lines. We use opacity sampling tables with a microturbulence parameter of $\xi=1.0$\,km\,s$^{-1}$ to approximate its contribution to spectral line broadening. The inaccuracies of this method should be small as we only investigate broadband-integrated spectra. Moreover, our wavelength resolution is only sufficient for a statistical representation of the spectra and does not resolve the profile shapes of most lines.

A representation of a stellar atmosphere with a plane-parallel stratification inherently limits the realism of surface brightness calculations near the limb, where a sharp drop-off should appear as soon as the line of sight no longer reaches below the atmosphere. \citet{Claretetal:2003} found such an intensity drop using 1D models with spherical symmetry. The position angle $\mu_{\mathrm{d}}$ of the drop depends on the stellar parameters, which influence the pressure scale height of the atmosphere, as well as on wavelength through the gas opacity. A simple estimate for $\mu_{\mathrm{d}}$ can be obtained by assuming a transparent atmosphere of one pressure scale height, which leads to the expression
\begin{equation}
\mu_{\mathrm{d}}=\sqrt{1-\left(\frac{R}{R+H}\right)},
\end{equation}
where $R$ is the stellar radius and $H$ is the pressure scale height. Adopting $R_{\mathrm{HD\,209458}}\approx1.23\,R_{\sun}$ and $R_{\mathrm{HD\,189733}}\approx0.77\,R_{\sun}$ from \citet{vanBelleetal:2009}, as well as choosing $H_{\mathrm{HD\,209458}}\approx200$\,km and $H_{\mathrm{HD\,189733}}\approx100$\,km at the surface as predicted by our 3D simulations, we obtain a drop-off at $\mu_{\mathrm{d}}\approx0.02$ for both stars. This result is in good agreement with \citet{Claretetal:2003}, who find $\mu_{\mathrm{d}}\approx0.05$ for a star with $T_{\mathrm{eff}}=5000$\,K and $\log g=4.0$. The lower surface gravity should yield a larger atmospheric pressure scale height compared to HD~189733, which has a very similar effective temperature. Our 3D models with plane-parallel stratification should therefore predict surface brightness distributions with sufficient realism for light curve analyses, except for some extreme cases of grazing transits.

\subsection{Analytical limb darkening laws}

\texttt{SCATE} yields monochromatic surface intensities $I_{\lambda}(x,y,\mu,\phi,t)$ at wavelength $\lambda$ for a range of discrete positions on the stellar disk, assuming plane-parallel geometry as the underlying 3D model atmospheres, with the projection factor $\mu=\cos\theta$ for angle $\theta$ in radial direction off the disk center. The resulting specific intensities are averaged over direction angle $\phi$, horizontal dimensions $x$ and $y$ at each angle, and time $t$:
\begin{equation}
\left<I_{\lambda}\right>_{x,y,\phi,t}(\mu)=\frac{1}{N_{t}}\sum_{N_{t}}\frac{1}{N_{\phi}}\sum_{N_{\phi}}\frac{1}{N_{x}N_{y}}\sum_{x,y}I_{\lambda}(x,y,\mu,\phi,t),
\end{equation}
where $N_{t}$, $N_{\phi}$, $N_{x}$, and $N_{y}$ denote the number of snapshots in the 3D time series, the number of azimuth angles, and the number of points on the horizontal axes. Limb darkening laws $I_{k}(\mu)/I_{k}(1)$ in a given wavelength band $k$ are then derived by numerically integrating the averaged surface intensities $\left<I_{\lambda}\right>(\mu)$ over the band:
\begin{equation}
\frac{I_{k}(\mu)}{I_{k}(1)}=\frac{\sum_{i=1}^{N_{\lambda}}\left<I_{\lambda_{i}}(\mu)\right>\Delta\lambda_{i}}{\sum_{i=1}^{N_{\lambda_{i}}}\left<I_{\lambda_{i}}(1)\right>\Delta\lambda_{i}}.
\label{eqn:lddef}
\end{equation}
The result is a smooth function of $\mu$. Interpolation of the limb darkening samples in $\mu$ using, e.g., cubic splines yields the best representation of the numerical results, but this method is more difficult to handle in the computation of model light curves. Sufficient accuracy can still be achieved with simple analytical fit formulae for stellar types that are relevant to planetary transit analyses, even though this approach may be more problematic for other cases \citep[see, e.g., the results for supergiant stars in][]{Chiavassaetal:2009}. The simple linear law,
\begin{equation}
\label{eqn:linearlaw}
\frac{I(\mu)}{I(1)}=1-u\left(1-\mu\right),
\end{equation}
is generally not a good approximation as it cannot capture the inherent nonlinearity of stellar limb darkening (see Fig.~\ref{fig:ldtestldphirescompare}), which may lead to unphysical results in the light curve analysis \citep{Southworth:2008}. We use a widely applied formula proposed by \citet{Claret:2000} for plane-parallel calculations,
\begin{equation}
\label{eqn:nonlinearlaw}
\frac{I(\mu)}{I(1)}=1-\sum_{n=1}^4c_{n}\left(1-\mu^\frac{n}{2}\right),
\end{equation}
which is fitted to the intensity points with equal weighting. The linear combination of 4 power laws exhibits some degeneracy in the coefficients, which can impede a per-coefficient comparison of limb darkening laws predicted by different models. However, the law is sufficiently versatile to fit a large range of brightness distributions with very small residuals and very good flux conservation (see Table~\ref{tab:ldtest}), which avoids the inaccuracies and ambiguities of using lower-order laws; see, e.g., the discussion in \citet{Heyrovsky:2007}.

\subsection{Numerical resolution of the 3D calculations}\label{sec:resolution}

The monochromatic surface brightness distribution on the stellar disk is resolved with $N_{\mu}=17$ non-equidistant points in the range $0.01\le\mu\le1.0$, adopting the same set of angles that is used for the Kurucz grid (see the left panel of Fig.~\ref{fig:ldtestldphirescompare}), and $N_{\phi}=4$ angles per point for all $\mu<1.0$. We reduce the horizontal resolution $N_{x}=N_{y}=240$ of the model atmospheres by half to accelerate the computation. The resulting intensities are averaged over time sequences with $N_{t}=10$ snapshots that span $\approx6$ periods of the fundamental pressure oscillation mode in each model. This is equivalent to $\approx70$\,min of stellar time in the case of HD~209458 and $\approx30$\,min for HD~189733.

We test the accuracy of our choice of numerical resolution by comparing cubic spline-interpolated theoretical limb darkening predictions in the spectral region between $3400$\,{\AA} and $3500$\,{\AA}, where limb darkening is strong and deviates significantly from a linear shape, using an arbitrary snapshot of the 3D model for HD~189733.

The numerical result for the computation with $N_{t}=1$, and with our ``standard'' resolution $N_{x}=N_{y}=120$, $N_{\phi}=4$ and $N_{\mu}=17$ is plotted as crosses in the left panel of Fig.~\ref{fig:ldtestldphirescompare}. The solid line shows a least-squares fit using the nonlinear law (Eq.~\ref{eqn:nonlinearlaw}), which yields a good representation of the model limb darkening. The dotted line shows a linear law (Eq.~\ref{eqn:linearlaw}), which clearly does not provide sufficient accuracy. The gray-shaded area above $\mu=0.86$ approximately indicates the inner region of the stellar disk that is never reached by the transiting planet, assuming an impact parameter $b=0.663$ and a planet-to-star radius ratio $R_{\mathrm{Planet}}/R_{\mathrm{Star}}=0.158$ (see Table~\ref{tab:hd189733fit}).

The right panel of Fig.~\ref{fig:ldtestldphirescompare} compares spline-interpolated computations with $N_{\phi}=8$ (solid line) and with the full horizontal resolution of the 3D models ($N_{x}=N_{y}=240$, dotted line), keeping the other parameters constant. The relative deviation from the standard resolution is smaller than $0.2$\,\% in either case, indicating that limb darkening is well-represented by our choices.

Larger sensitivity of the interpolated limb darkening law is observed when the number of $\mu$ angles is increased, refining the radial sampling of the surface brightness distribution. The left panel in Fig.~\ref{fig:ldmucompare} shows the relative deviation of interpolated computations with $N_{\mu}=17$ points (solid line) and $N_{\mu}=51$ (dotted line) from $N_{\mu}=101$. The node distribution of the $N_{\mu}=51$ and $N_{\mu}=101$ cases follows Gauss-Legendre quadrature (plus disk center at $\mu=1.0$). The $N_{\mu}=17$ set reaches 1\,\% relative deviation near the limb at $\mu=0.01$ as the interpolated curve is visibly less constrained between the nodes (marked by grey vertical lines in the plot). The differences above $\mu\gtrsim0.1$ are practically negligible, as are those between the $N_{\mu}=51$ and the $N_{\mu}=101$ cases (dotted line). The nonlinear law cannot represent such small-scale variations in the limb darkening curve; increasing $N_{\mu}$ leads to slightly larger deviation of the computed intensity samples from the best-fit curve (right panel of Fig.~\ref{fig:ldmucompare}), while the curve itself changes weakly. Table~\ref{tab:ldtest} compares the best-fit coefficients of the nonlinear law (Eq.~\ref{eqn:nonlinearlaw}) and the reduced $\chi^{2}$ that results from the different choices of $N_{\mu}$. The degeneracy of the four coefficients does not allow a direct comparison; they change by up to $\approx30$\,\% ($c_{4}$) between $N_{\mu}=17$ and $N_{\mu}=101$, which is much larger than the actual point-to-point difference in the curves. The inability of the nonlinear law to take small-scale variations into account is reflected by an increasing $\chi^{2}_{\nu}$ for increasing $N_{\mu}$.

Table~\ref{tab:ldtest} also compares the surface flux predicted by the $\mu$-integral of the nonlinear law,
\begin{equation}
\frac{F_{\mathrm{nl}}}{\pi I(1)}=\frac{1}{\pi}\int_{\phi=0}^{2\pi}\int_{\mu=0}^{1}\frac{I(\mu)}{I(1)}\mu d\mu d\phi=1-\sum_{n=1}^{4}c_{n}\left(1-\frac{4}{n+4}\right),
\end{equation}
with the direct numerical integral $F_{\mathrm{num}}$ of the 3D computation, which uses the trapezoid rule for $N_{\mu}=17$ and Gauss-Legendre quadrature for $N_{\mu}=51$ and $N_{\mu}=101$ (the disk center has zero weight in the latter two cases). We obtain excellent flux conservation for the nonlinear law with a relative deviation of less than $10^{-4}$ for $N_{\mu}=51$ and $N_{\mu}=101$. The trapezoid rule with $N_{\mu}=17$ leads to less accurate integration, as the sampling of the surface brightness distribution is less optimal compared to Gauss-Legendre quadrature. The integral of the fitted nonlinear law, however, deviates much less from the computations with $N_{\mu}=51$ and $N_{\mu}=101$, as the fit appears less sensitive to the sampling of the surface brightness distribution, which emphasizes again that the number of $\mu$ angles in our standard resolution is sufficient.

Our full spectrum computations contain surface intensities for a set of $\approx108000$ wavelengths over a spectral range between $910$\,{\AA} and $20$\,$\mu m$ with a constant sampling rate of $\lambda/\Delta\lambda=20000$. This statistical treatment of the stellar spectrum leads to inaccuracies in the spectral energy distribution (SED) when narrow wavelength ranges are investigated due to missing information on spectral line profiles between the opacity samples. \citet{Plez:2008} estimated the uncertainty of the solar radiative flux derived from a \texttt{MARCS} model for constant wavelength bin sizes of $\Delta\lambda_{i}=\lambda_{i}/200$, where $\lambda_{i}$ is the center wavelength in bin $i$, by reducing the number of opacity samples per bin by factors of 3. Errors reach the $5\,\%/\sqrt{3}\approx3\,\%$ level at $\lambda=4000$\,{\AA} with $\Delta\lambda=20$\,{\AA}, they increase towards the UV and decrease towards longer wavelengths.

The wavelength sampling rate affects the strength of band-integrated limb darkening, which is weaker in strong lines than at continuum wavelengths (see the discussion in Appendix~\ref{sec:ldtttau}). More frequent and stronger lines in the opacity sampling thus effectively lead to weaker limb darkening. We test the importance of wavelength resolution between $3400$\,{\AA} and $3500$\,{\AA} by reducing the number of sampling points by 2 and 4, resulting in sampling rates of $R=10000$ and $R=5000$, respectively. The relative deviations of spline-interpolated limb darkening predictions from full wavelength resolution are compared in the left panel of Fig.~\ref{fig:ldossequencecompare}, where solid lines show the two possible choices of resampling for $R=10000$ and dotted lines show the four possible choices for $R=5000$. The deviation can reach up to $4.5$\,\% near the limb at the lowest sampling rate. It is important to keep in mind that this result strongly depends on the width of the wavelength band for which limb darkening is calculated and on the wavelength region itself through the presence or absence of spectral lines.

We finally investigate the influence of time-dependent variation in the atmospheric stratification. Similar to real stars, 3D time-dependent hydrodynamical models exhibit pressure-mode oscillations that lead to periodic steepening of the mean $T$-$\tau$ gradient and, as a consequence, to slight variations in the strength of limb darkening (see also Sect.~\ref{sec:3D1Dlimbdark}). Surface intensities are averaged over a sample of $N_{t}=10$ snapshots in our standard setting in order to obtain a robust representation, typically covering $\sim6$ periods of the fundamental oscillation mode. We compare the resulting limb darkening predictions with a time sequence that is twice as long (solid line in the right panel of Fig.~\ref{fig:ldossequencecompare}) and find a relative deviation of up to $1.5$\,\% near the limb, declining quickly towards the disk center. The difference between the two sets is slightly larger (dotted line).

\subsection{Summary}

Most uncertainties of numerical resolution affect the limb darkening predictions only very close to the limb beneath $\mu\lesssim0.05$, a region where the assumption of plane-parallel geometry starts to yield generally unrealistic radiative intensities as it was discussed above. This affects the earliest ingress and latest egress phases during the transit of HD~209458 and HD~189733. Characteristic residuals in the model light curve that result from inaccurate predictions near the limb are not observed (see Sect.~\ref{sec:lcfits}), although the available data only covers parts of the critical regions and may not be sufficiently accurate to clearly highlight discrepancies. There is still some influence of the calculations at small $\mu$ on the entire limb darkening law due to the finer sampling, giving it more relative weight in the least-squares fit. As HST light curve analyses based on our 3D model for HD~189733 were shown to yield excellent results that rival the quality of direct fits of limb darkening \citep[see Table~2 in][]{Singetal:2011}, we trust that our calculations provide a sufficiently accurate and realistic description of the effect.

The assumptions and simplifications used in this work may fail in other cases: time-variation of the atmospheric temperature gradient may be a significant issue for very weak transit signals which occur in systems with smaller planet-to-star radius ratios, such as earth-sized planets with solar-type host stars. The effects of spherical geometry will certainly become more important for transiting planets that reach only the outermost parts of the stellar disk. We also stress that an investigation of limb darkening for increasingly narrow wavelength bands requires more detailed spectrum synthesis with better wavelength resolution, as the wavelength sampling effects will become increasingly important. This issue affects observations with future generations of telescopes that will allow narrowband transit spectroscopy of exoplanetary atmospheres.

\section{Limb darkening predictions of the 3D models and 1D models}\label{sec:3D1Dlimbdark}

\begin{figure*}[htbp]
\centering
\includegraphics[width=8.5cm]{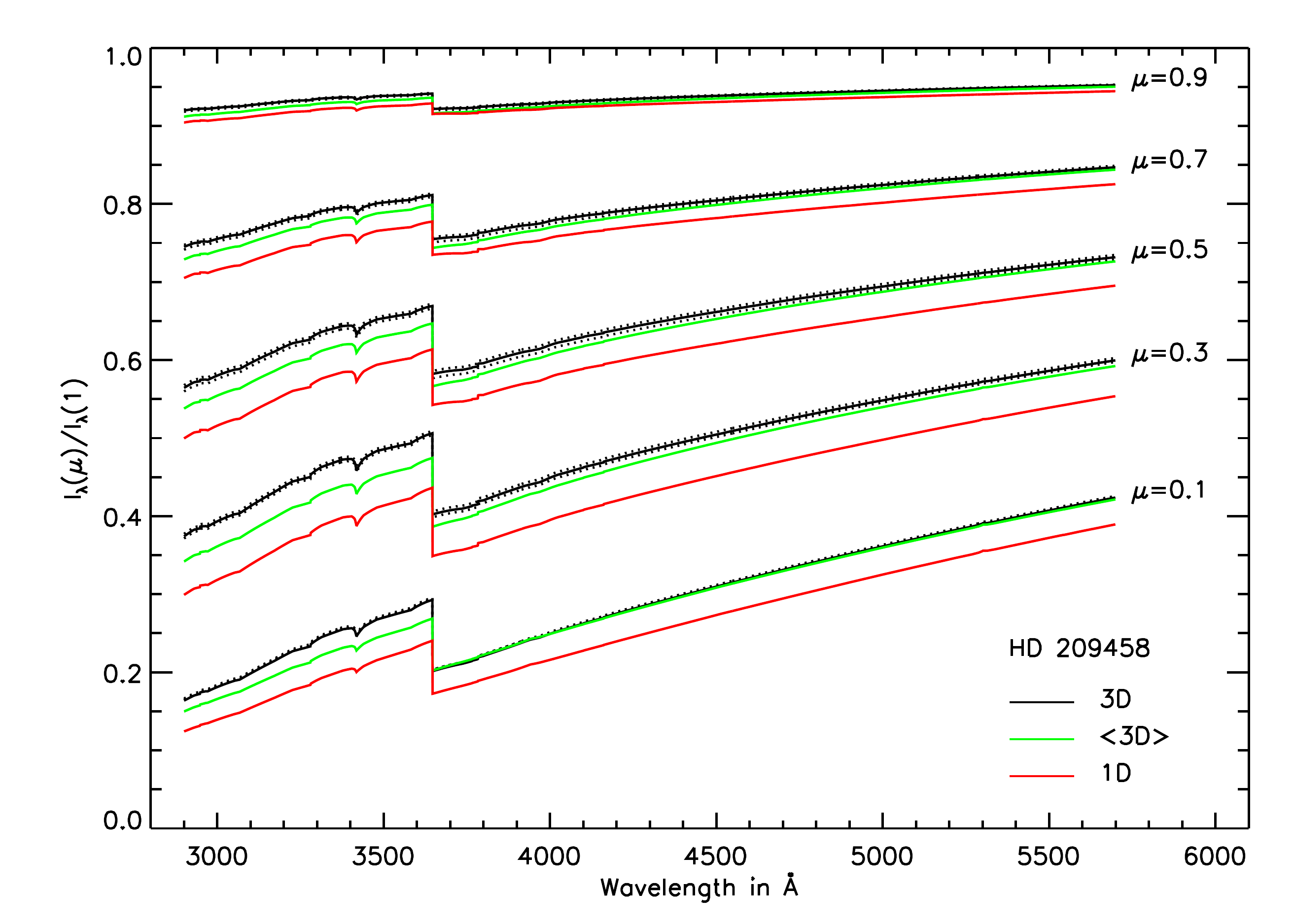}
\includegraphics[width=8.5cm]{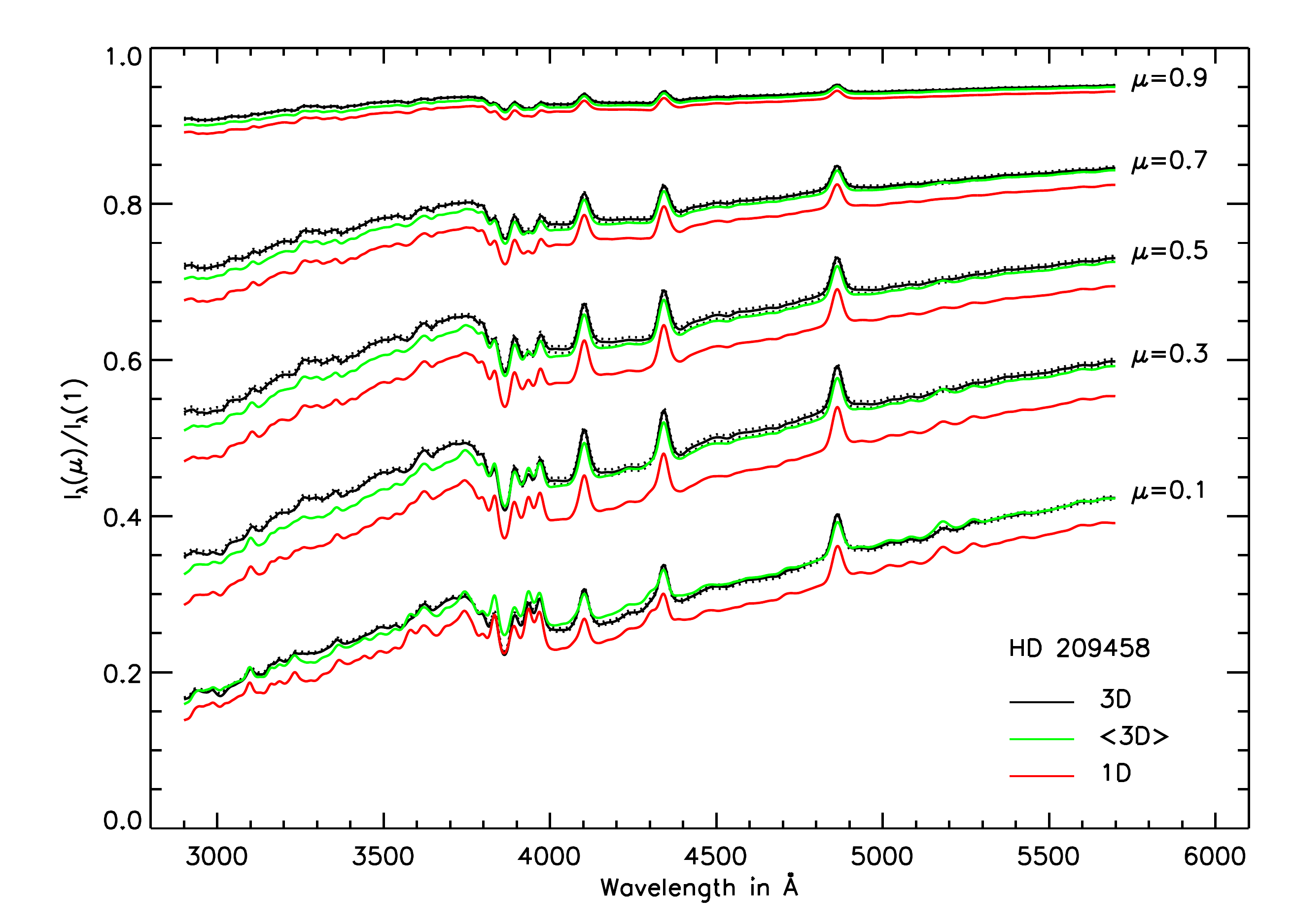}
\caption{Surface intensities $I_{\lambda}(\mu)/I_{\lambda}(1)$ for HD~209458 at different $\mu$ angles, including only continuous absorption (left panel) as well as continuous and spectral line absorption (right panel) in the radiative transfer computation. Time-averaged 3D spectra (black solid lines) are compared to spectra derived from the mean 3D model (green solid lines) and the 1D \texttt{MARCS} model (red solid lines). Dotted black lines show spectra derived from 3D snapshots that exhibit intensity minima and maxima at $5700$\,{\AA}, bracketing the time variation of the 3D surface brightness due to oscillations in the model atmosphere. Note that spectra in the right panel were convolved with a Gaussian kernel with $\lambda/\Delta\lambda=250$ for clarity.}
\label{fig:ldcontlineshd209458}
\end{figure*}

\begin{figure*}[htbp]
\centering
\includegraphics[width=8.5cm]{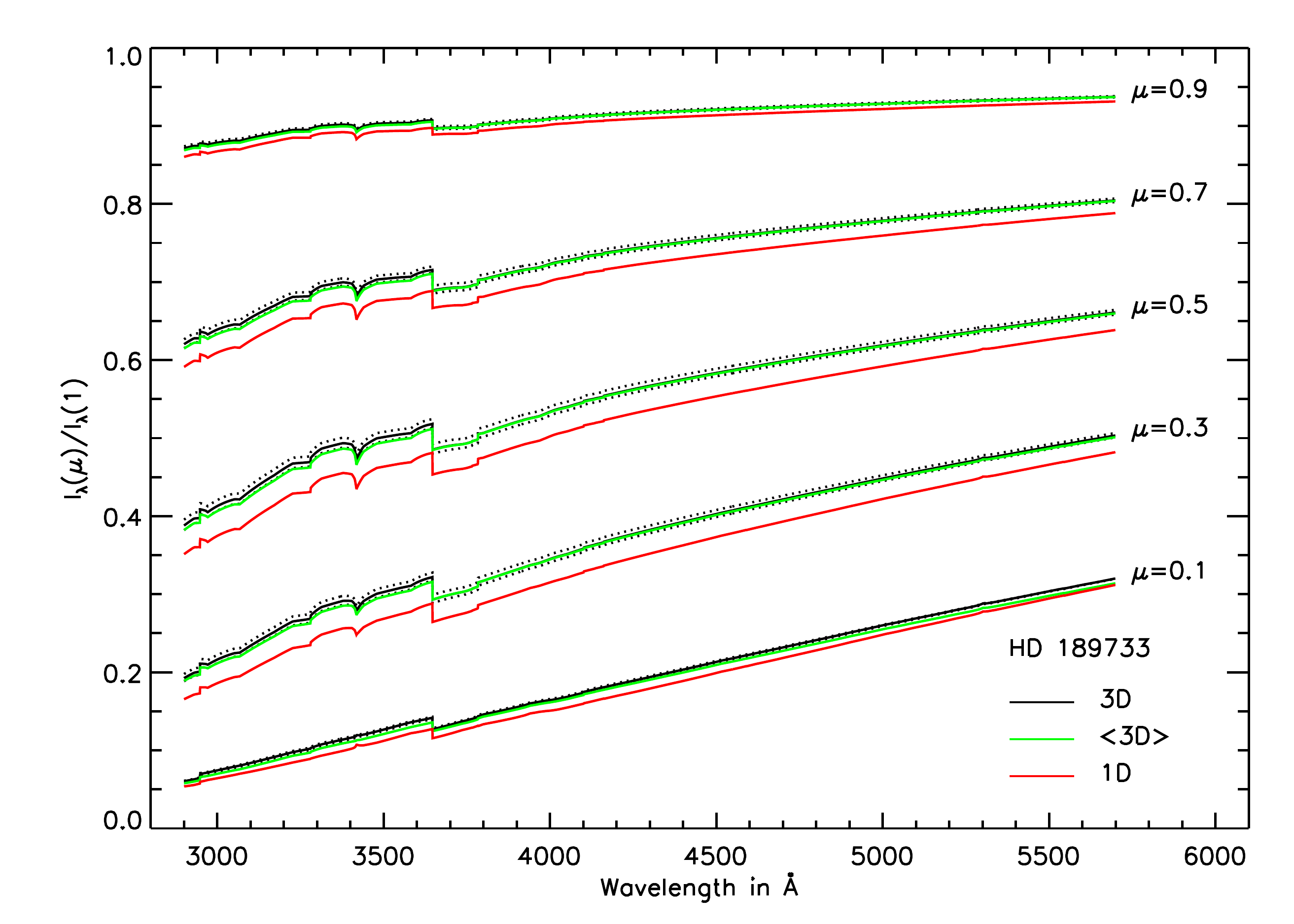}
\includegraphics[width=8.5cm]{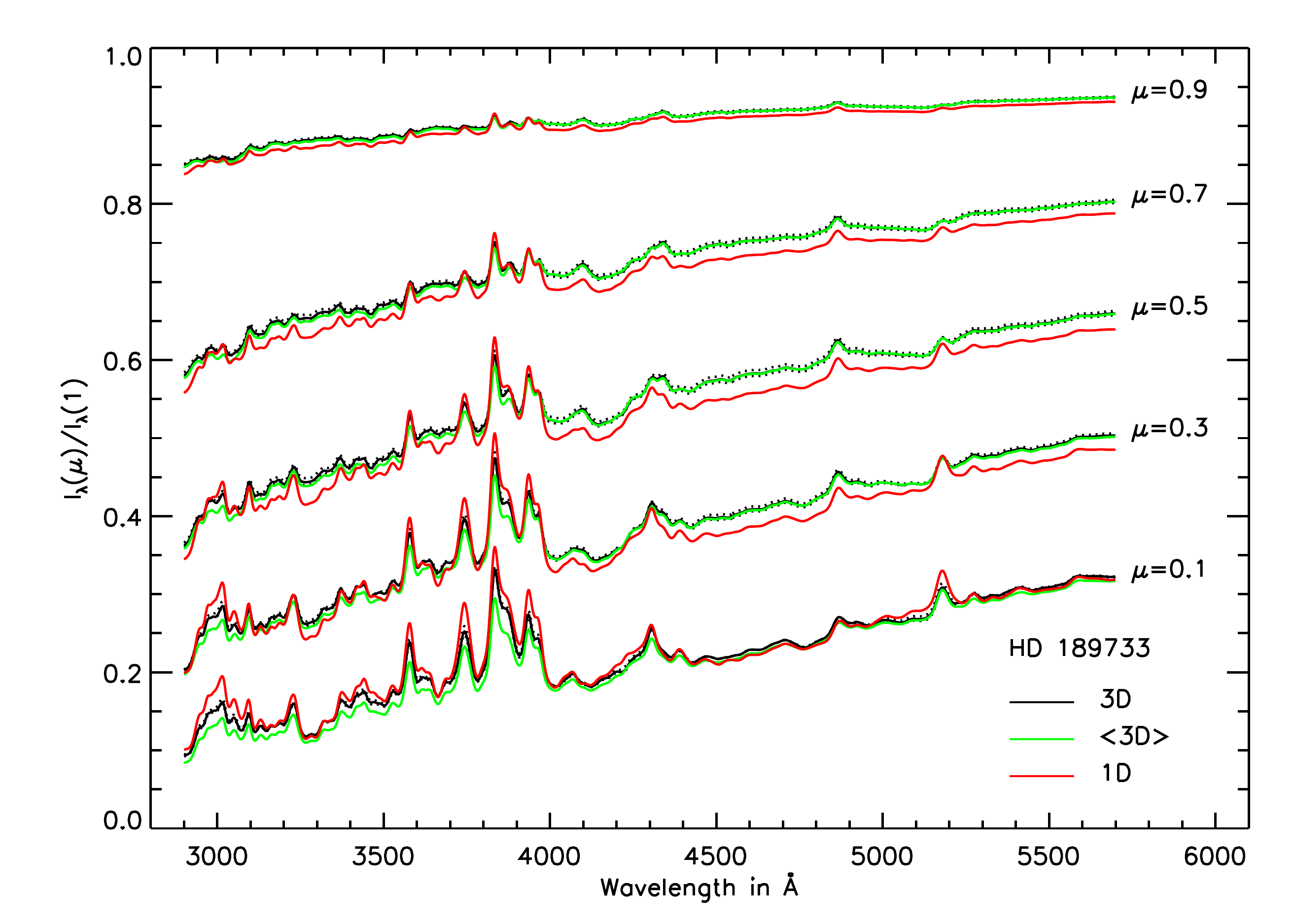}
\caption{Same as Fig.~\ref{fig:ldcontlineshd209458}, computed for HD~189733.}
\label{fig:ldcontlineshd189733}
\end{figure*}

\begin{figure*}[htbp]
\centering
\includegraphics[width=8.5cm]{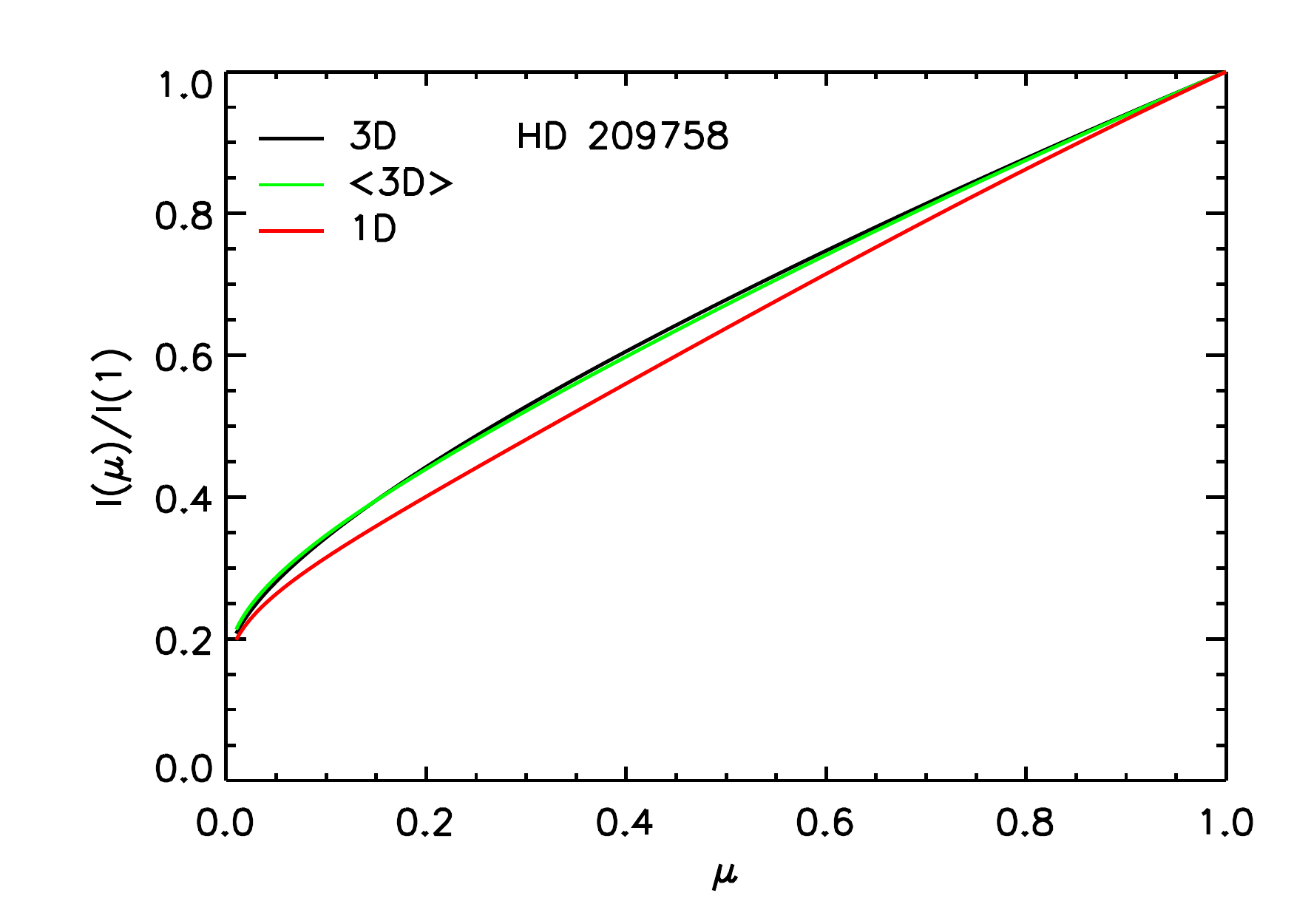}
\includegraphics[width=8.5cm]{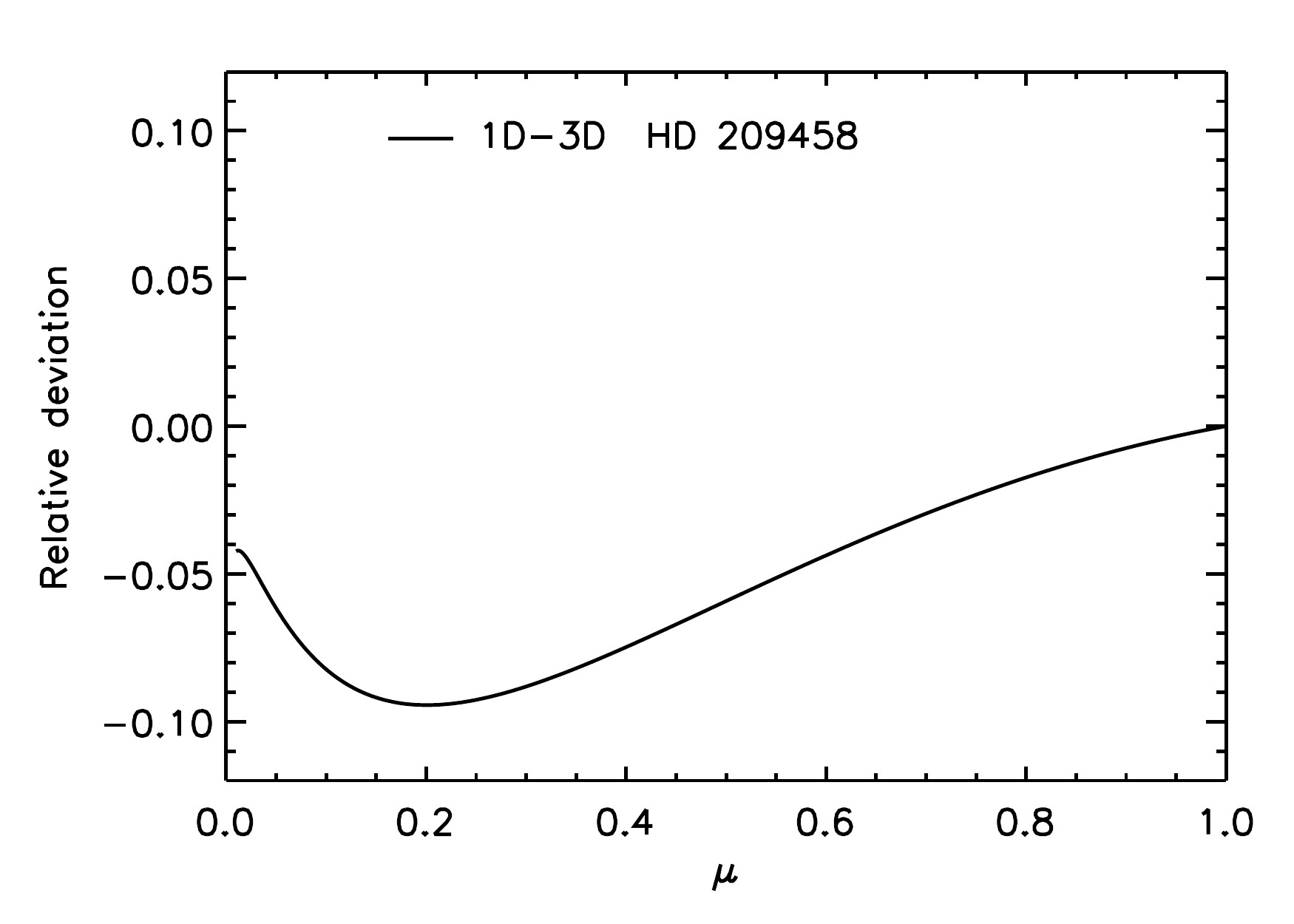}
\caption{\textit{Left:} Wavelength-integrated limb darkening curves for the 3D model of HD~209458 (black solid line), the mean 3D model (green solid line) and the 1D \texttt{MARCS} model (red solid line). Surface intensities between $2900$\,{\AA} and $5700$\,{\AA} were weighted with the sensitivity function of the HST STIS instrument and wavelength (see Appendix~\ref{sec:ldccompilation}). \textit{Right:} Relative deviation between the limb darkening predictions of the 1D \texttt{MARCS} model with respect to the 3D model.}
\label{fig:ld3d1dwhitelight209458}
\end{figure*}

\begin{figure*}[htbp]
\centering
\includegraphics[width=8.5cm]{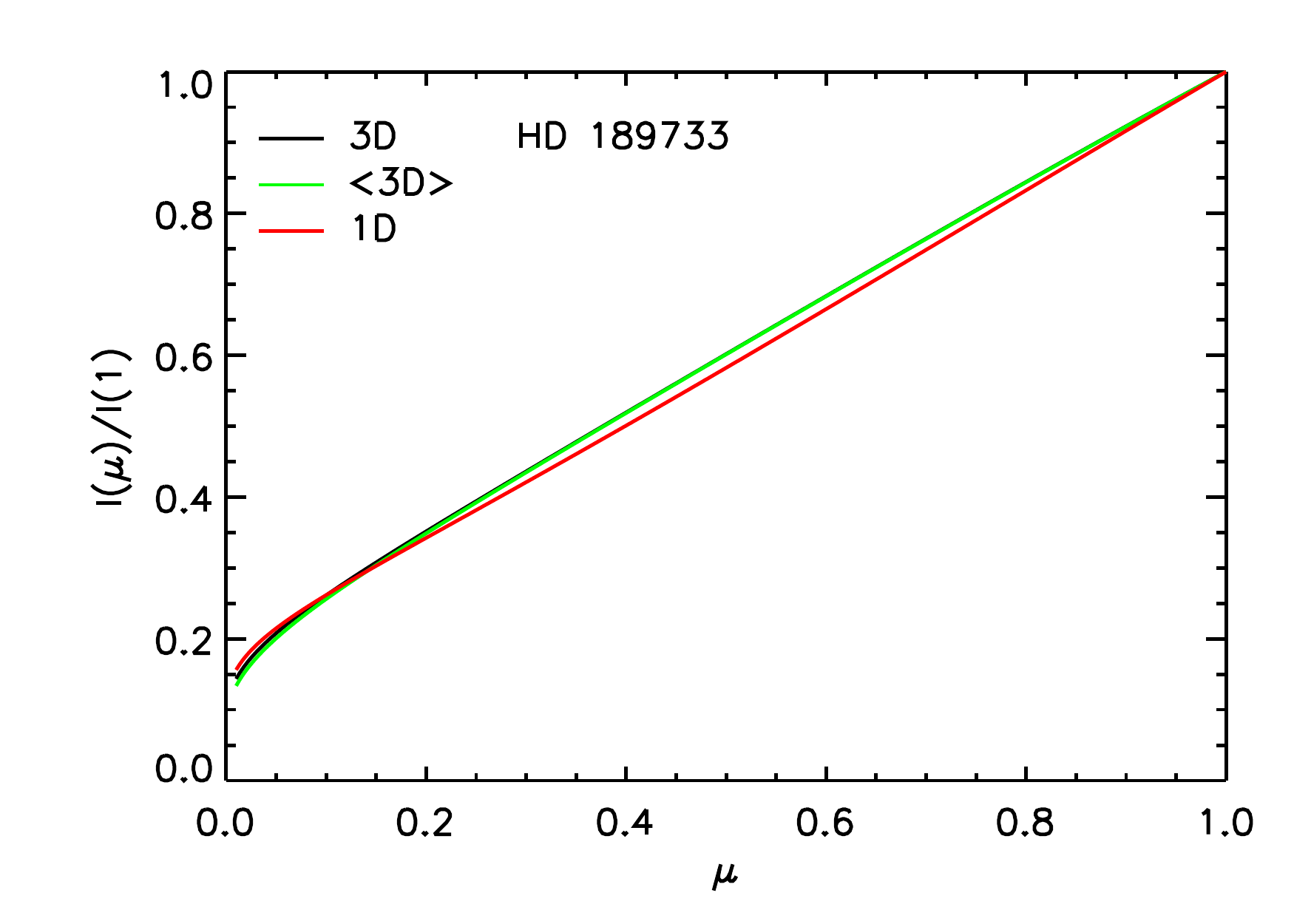}
\includegraphics[width=8.5cm]{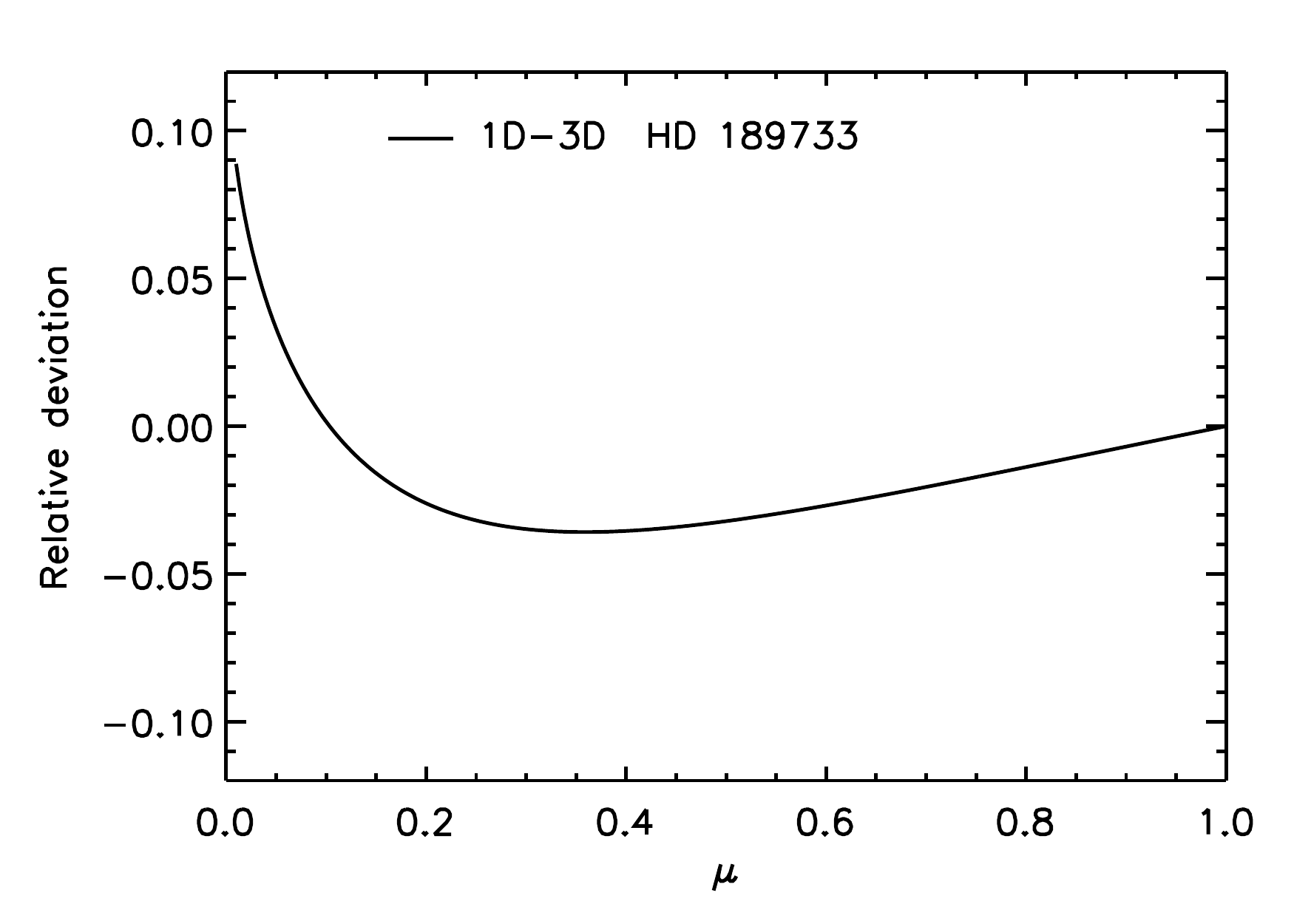}
\caption{Same as Fig.~\ref{fig:ld3d1dwhitelight209458} for the case of HD~189733.}
\label{fig:ld3d1dwhitelight189733}
\end{figure*}

Classical 1D models of the solar atmosphere are known to overestimate the effect of limb darkening compared to solar continuum observations of \citet{Neckeletal:1994}; \citet{Singetal:2008} examine an \texttt{ATLAS} model and \citet{Asplundetal:2009} a \texttt{MARCS} model. The latest generation of 3D hydrodynamical model atmospheres, however, exhibits very good agreement with the solar continuum \citep[see Fig.~2 in][]{Asplundetal:2009} and other observational tests \citep[e.g.][]{Pereiraetal:2009b,Pereiraetal:2009a}.

3D models take surface granulation explicitly into account rather than parameterizing convection with a simplified recipe, producing strong horizontal temperature fluctuations and time-dependent pressure oscillations that affect the temperature structure. The spatial and temporal mean temperature gradient thus deviates from a 1D model with the same stellar parameters. While this often accounts for the leading order effect, the temperature fluctuations with respect to the mean stratification can also have an impact on the emitted radiation.

Quantitative limb darkening predictions from numerical models rely on a complex nonlinear coupling of radiative transfer, plasma dynamics and microphysics. The main mechanisms can still be understood using basic results from 1D stellar atmosphere theory, which yields the wavelength-dependent approximate relation
\begin{equation}
\label{eqn:ldeddingtonbarbier}
u_{\lambda}\approx\frac{\log e}{B_{\lambda}(T(\tau_{\lambda}=1))}\left.\frac{dB_{\lambda}}{dT}\right|_{\tau_{\lambda}=1}\left.\frac{dT}{d\log\tau_{\lambda}}\right|_{\tau_{\lambda}=1}
\label{eqn:ldapprox}
\end{equation}
for the steepness $u_{\lambda}$ of a linear limb darkening law (cf. Eq.~(\ref{eqn:linearlaw})). $B_{\lambda}$ is the Planck function and $T(\tau_{\lambda})$ is the temperature stratification of the model atmosphere as a function of monochromatic vertical optical depth $\tau_{\lambda}$; see Appendix~\ref{sec:ldtttau} for a derivation. In this linear model, the strength of limb darkening at a given wavelength scales directly with the temperature gradient at the optical surface $\tau_{\lambda}=1$. The wavelength-dependence of limb darkening stems from the relative temperature sensitivity of the Planck function, expressed through the term $(dB_{\lambda}/dT)/B_{\lambda}$, which generally leads to an increase towards shorter wavelengths, as well as from opacity variations that change the monochromatic optical depth. The latter effect can play an important role in spectral features, such as strong absorption lines, and leads to weaker limb darkening if the temperature decreases monotonously outward; see Appendix~\ref{sec:ldtttau} for a discussion. The simple model behind Eq.~(\ref{eqn:ldapprox}) assumes linearity of the Planck function with optical depth and thus cannot include any variation of the gradients found in real stellar atmospheres; its applicability is therefore limited to a more or less narrow region around the optical surface.

\subsection{HD~209458}

We compute surface intensities in the spectral range between $2900$\,{\AA} and $5700$\,{\AA} for the 3D model of HD~209458, its spatially and temporally averaged 1D stratification (mean 3D model), and the 1D \texttt{MARCS} model which has a lower effective temperature (see Table~\ref{tab:stellp}). The left panel in Fig.~\ref{fig:ldcontlineshd209458} compares continuum intensities as a function of wavelength and $\mu$ for the different models.

The wavelength-dependence of limb darkening is clearly visible in the decreasing relative intensity towards shorter wavelengths for given $\mu$, resulting in a stronger effect. The opacity-dependence of limb darkening manifests itself in the reverse Balmer jump at $3646$\,{\AA}: on the blue side, the additional \ion{H}{I} bound-free opacity leads to weaker limb darkening compared to the red side where the atmosphere is more transparent.

The 3D model (black solid lines in Fig.~\ref{fig:ldcontlineshd209458}, the spectrum of each individual 3D snapshot is computed individually and then averaged over time) predicts significantly weaker limb darkening than the 1D \texttt{MARCS} model (red solid lines) in the $\mu$ range shown in the plot. The deviations decrease towards the limb as the 3D and 1D temperature gradients in the higher atmosphere become more similar. Using a 1D \texttt{MARCS} model with the same effective temperature as the 3D model (not shown in the plot) only leads to a small deviation from the cooler 1D model. The result of the 3D-1D comparison is similar to the solar case \citep[see Fig.~2 in][]{Asplundetal:2009}, as expected from the similarity of stellar parameters. 

The mean 3D model (green sold lines in Fig.~\ref{fig:ldcontlineshd209458}) yields almost the same continuous spectrum as the full 3D model well down to the Balmer jump, indicating that the mean temperature stratification dominates the 3D effect in this part of the spectrum. At shorter wavelengths, temperature fluctuations play a greater role for continuum formation, leading to larger deviations between the 3D prediction and the mean 3D result. The time-dependent pressure oscillations and temperature fluctuations of the 3D model lead to slight variation in the normalized intensities; the black dotted lines indicate the spectra with minimum and maximum intensity at $5700$\,{\AA}.

Strong absorption features appear inverted in the right panel of Fig.~\ref{fig:ldcontlineshd209458} due to weaker limb darkening for larger opacity. Broadband limb darkening curves therefore integrate a variety of limb darkening strengths if spectral lines are present, depending on number and strength of the features. It is thus essential to use the same line data and chemical composition for a model comparison as it is performed in this paper. The deviation between 3D spectra and 1D spectra decreases in the presence of line absorption: a larger atmospheric height range influences the outgoing radiation field, with temperature gradients of the 3D model and 1D \texttt{MARCS} model becoming more similar higher up in the atmosphere. The slightly hotter temperatures of the 1D \texttt{MARCS} model beneath $\log\tau_{5000}\lesssim-2$ remove or even reverse the 3D-1D difference; the effect is visible in the features around $3800$\,{\AA} in the right panel of Fig.~\ref{fig:ldcontlineshd209458}.

Wavelength-integrated limb darkening curves for HD~209458 are shown in the left panel of Fig.~\ref{fig:ld3d1dwhitelight209458}, the monochromatic intensity distributions between $2900$\,{\AA} and $5700$\,{\AA} were weighted with the sensitivity function of the STIS instrument and with wavelength (see Appendix~\ref{sec:ldccompilation}). The shape of the 3D curve is nonlinear across the stellar disk, with a slightly steeper drop near the limb. A comparison of the 3D and 1D \texttt{MARCS} results immediately reveals a clear distinction between the two models, despite the influence of spectral lines. The mean 3D model produces almost the same limb darkening curve as the full 3D model. The right panel of Fig.~\ref{fig:ld3d1dwhitelight209458} shows the relative deviation between the 3D and 1D results, which decreases close to the limb as the temperature stratification of the 1D \texttt{MARCS} model becomes shallower compared to the 3D model in the higher atmosphere.

\subsection{HD~189733}

The same computation is repeated for HD~189733; the left panel of Fig.~\ref{fig:ldcontlineshd189733} shows continuum intensities, the right panel continuum and line intensities. The 3D model (black solid lines) again predicts overall weaker limb darkening in the continuum, but the difference between the 3D and 1D \texttt{MARCS} (red solid lines) results is visibly smaller compared to HD~209458, owing to the smaller difference in the atmospheric temperature stratifications of the models deeper in the photosphere (see Fig.~\ref{fig:3Dmodel}). The mean 3D model (green solid lines) represents the continuous spectrum well also below the Balmer jump, indicating that the mean temperature gradient of the 3D model has the dominant influence. The time variation of the 3D curves, indicated by black dotted lines in Fig.~\ref{fig:ldcontlineshd189733}, is small.

The 1D \texttt{MARCS} model is again warmer than the 3D model in the higher atmosphere, but the difference is much more pronounced for HD~189733. The shallow 1D temperature gradient leads to significantly smaller deviations between the 3D and 1D continuum intensities at $\mu=0.1$ compared to the region around, e.g., $\mu=0.5$. The consequences of the hotter 1D temperatures become even more apparent when spectral lines are included in the computation (right panel of Fig.~\ref{fig:ldcontlineshd189733}). The cores of the strongest lines that form in the higher photosphere exhibit weaker limb darkening in 1D than in 3D, visible in the features beneath $4000$\,{\AA}.

Spectral line absorption is overall stronger in HD~189733 due to its cooler atmospheric temperatures. The broadband-integrated limb darkening curves therefore include a larger variety of limb darkening strengths and are more heavily influenced by temperature gradients at different atmospheric heights. The deviation between 3D and 1D limb darkening curves (black and red solid line in the left panel of Fig.~\ref{fig:ld3d1dwhitelight189733}) is significantly smaller compared to HD~209458; the differences in the atmospheric temperature stratification are partially hidden through the influence of spectral lines. The mean 3D model (green solid line) predicts practically the same limb darkening curve as the full 3D model, with the exception of the region closest to the limb. The right panel of Fig.~\ref{fig:ld3d1dwhitelight189733} shows again the relative deviation between the 3D and 1D limb darkening curves. The hotter atmospheric temperatures of the 1D \texttt{MARCS} model lead to visibly weaker broadband limb darkening compared to the 3D model beneath $\mu\lesssim0.1$.

\section{Transit light curve fits}\label{sec:lcfits}

We test the theoretical limb darkening predictions of the 3D models and the 1D \texttt{MARCS} models using HST light curves for \object{HD~209458b} and \object{HD~189733b}. Observations from the Space Telescope Imaging Spectrograph (STIS) with the G430L grating were integrated over the wavelength range between $2900$\,{\AA} and $5700$\,{\AA} where limb darkening is strong; theoretical intensity spectra were weighted with the sensitivity function of the instrument (see Appendix~\ref{sec:ldccompilation}; limb darkening coefficients for other instruments and a range of standard broadband filters are listed in Table~\ref{tab:ldccompilation}). Using large spectral bandwidth in the comparison maximizes the signal-to-noise ratio of the observations and reduces the uncertainties of the opacity sampling technique (Sect.~\ref{sec:resolution}). The differences between the atmospheric stratifications of 3D models and 1D models can be obscured by such broadband integration if spectral lines are present, as discussed in Sect.~\ref{sec:3D1Dlimbdark}, but light curve fits in smaller wavelength bands yielded essentially the same results for the 3D-1D comparison. To complement the sparse transit coverage of our HST STIS data for HD~189733, we also include light curve fits of observations between $5350$\,{\AA} and $10500$\,{\AA} that were obtained with the HST Advanced Camera for Surveys (ACS) and the HRC G800L grating, although the limb darkening effect is weaker towards longer wavelengths.

Observed light curves are analyzed with the models of \citet{Mandeletal:2002} and using least-squares fits with the Levenberg-Marquardt algorithm. The ratio of orbital to stellar radius, the orbital inclination and the stellar density are well known for both systems and therefore fixed to their literature values. The central transit time needs to be fitted for HD~209458b and the ACS data for HD~189733b. The remaining free parameters are the planet-to-star radius ratio, stellar baseline flux and several polynomial coefficients that correct for the above mentioned systematic effects of the instrument \citep[see][]{Singetal:2011}.

Limb darkening affects the entire transit of the planet; its generally growing strength towards shorter wavelengths leads to an increasingly round shape of the light curve \citep[see Fig.~3 in][]{Knutsonetal:2007}. This causes some degeneracy between limb darkening and the planet-to-star radius ratio. The latter parameter thus varies slightly between a direct fit of limb darkening and using model predictions. In addition to that, the removal of instrument systematics from the observed stellar flux with a polynomial function is partly degenerate with the shape of the transit light curve, allowing for compensation of inaccurate limb darkening predictions in a simultaneous fit. We therefore first remove the systematics using the out-of-transit points in each visit and then fit the resulting light curves with limb darkening predictions from 3D and 1D models, keeping only the baseline flux and the planet-to-star radius ratio as free parameters in each case. While this split in the process does not necessarily produce the best fit to the data, it allows for a clear comparison of limb darkening predictions without any obliteration by other free parameters.

\subsection{HD~209458b}

\begin{table*}[htbp]
\caption{Parameters for the HST transit light curves of HD~209458b, shown in Fig.~\ref{fig:hd209458whitelight}.}
\begin{center}
\begin{tabular}{lllll}
\hline\hline
\multicolumn{5}{c}{HD~209458}\T\BB\\
\multicolumn{1}{c}{} & \multicolumn{2}{c}{3D} & \multicolumn{2}{c}{1D}\\
\hline
Parameter & $\phantom{-}$Visit 1 & $\phantom{-}$Visit 2 & $\phantom{-}$Visit 1 & $\phantom{-}$Visit 2 \T \\
\hline
$c_{1}$ & $\phantom{-}0.4755$ & $\phantom{-}0.4755$ & $\phantom{-}0.6266$ & $\phantom{-}0.6266$\T\\
$c_{2}$ & $\phantom{-}0.3451$ & $\phantom{-}0.3451$ & $                  - 0.6911$ & $                   -  0.6911$\\
$c_{3}$ & $\phantom{-}0.0719$ & $\phantom{-}0.0719$ & $\phantom{-}1.6275$ & $\phantom{-}1.6275$\\
$c_{4}$ & $                   -0.0488$ & $                    -0.0488$ & $                   - 0.7043$ & $                  -  0.7043$\\
\hline
Incl. [deg] & $\phantom{-}86.664$ & $\phantom{-}86.664$ & $\phantom{-}86.664$ & $\phantom{-}86.664$\T\\
$\rho_{\mathrm{Star}}$\,[g\,cm$^{-3}$] & $\phantom{-}1.036$ & $\phantom{-}1.036$ & $\phantom{-}1.036$ & $\phantom{-}1.036$ \\
$a_{\mathrm{Orbit}}/R_{\mathrm{Star}}$ & $\phantom{-}8.7947$ & $\phantom{-}8.7947$ & $\phantom{-}8.7947$ & $\phantom{-}8.7947$ \\
\hline
$F_{\mathrm{Baseline}}$\,$[10^{8}\,\mathrm{counts}]$ & $\phantom{-}1.391\phantom{00}\pm0.00004$ & $\phantom{-}1.391\phantom{00}\pm0.00004$ & $\phantom{-}1.391\phantom{00}\pm0.00044$ & $\phantom{-}1.391\phantom{00}\pm0.00049$\T \\ 
$R_{\mathrm{Planet}}/R_{\mathrm{Star}}$ & $\phantom{-}0.12156\pm0.00016$ & $\phantom{-}0.12164\pm0.00017$ & $\phantom{-}0.12121\pm0.00194$ & $\phantom{-}0.12140\pm0.00215$ \\
\hline
DOF\tablefootmark{1} & $\phantom{-}253$ & $\phantom{-}248$ & $\phantom{-}253$ & $\phantom{-}248$\T \\
$\chi^{2}$ & $\phantom{-}265$ & $\phantom{-}248$ & $\phantom{-}813$ & $\phantom{-}735$ \\
$\chi^{2}_{\nu}$ & $\phantom{-}1.05$ & $\phantom{-}1.00$ & $\phantom{-}3.21$ & $\phantom{-}2.96$\BB \\
\hline
\end{tabular}
\tablefoot{
Fitted parameters have error bars. Parameters kept constant for different fits appear repeatedly to improve readability of the table.\\
\tablefoottext{1}{Degrees of freedom.}
}
\end{center}
\label{tab:hd209458fit}
\end{table*}

\begin{figure*}[htbp]
\centering
\includegraphics[width=8.5cm]{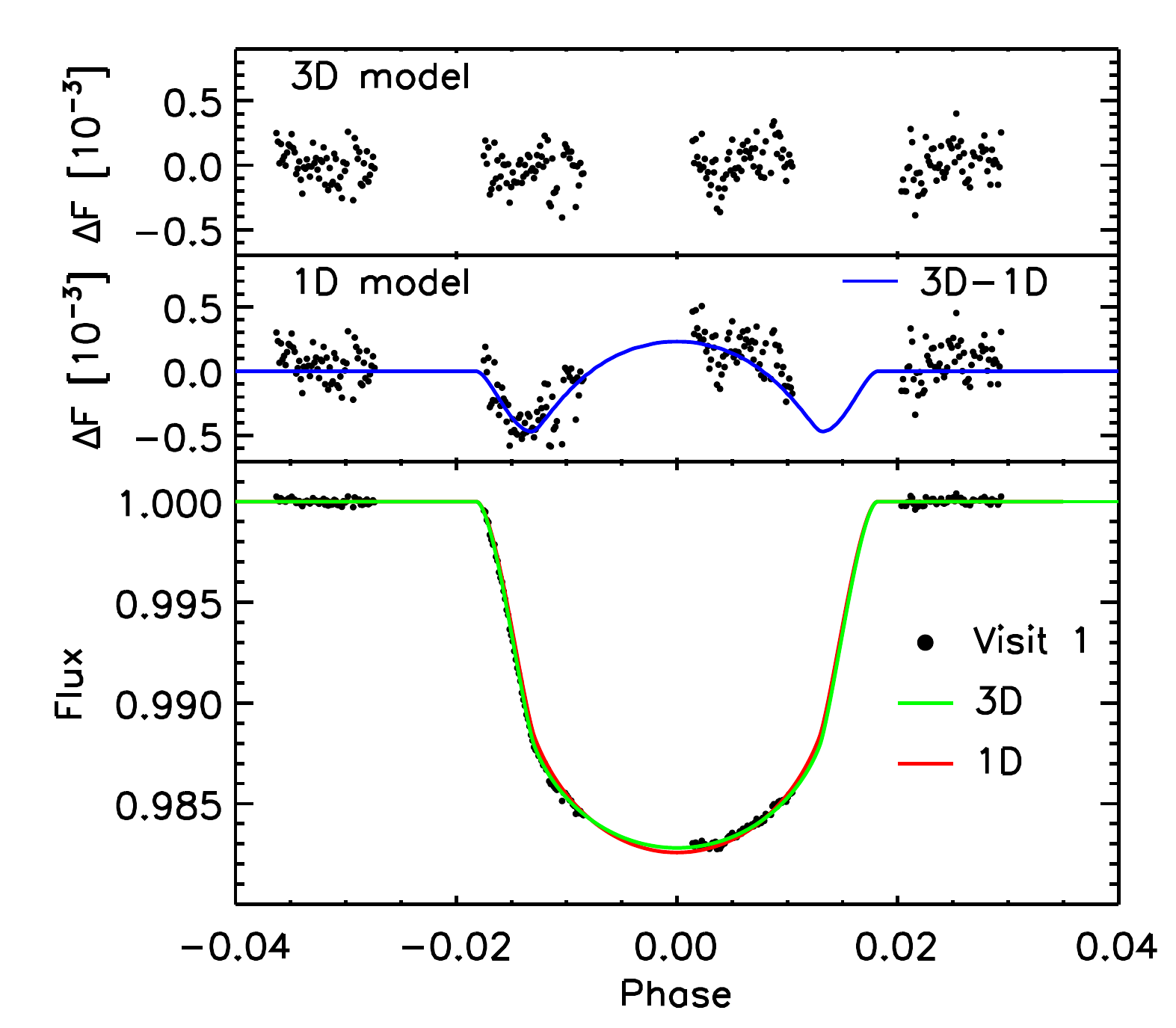}
\includegraphics[width=8.5cm]{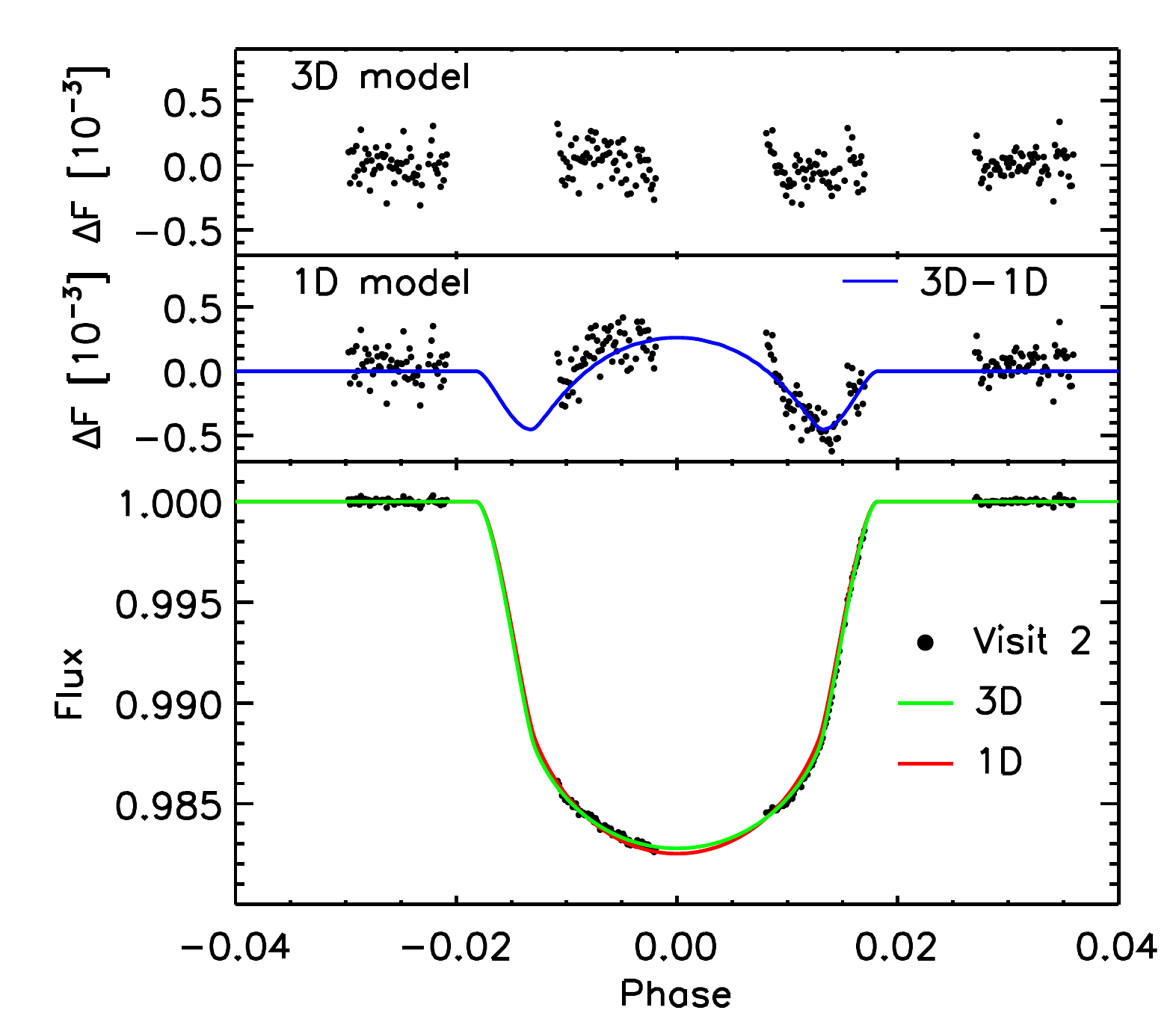}
\caption{Transit light curve fits for two visits of HD~209458b (dots in the lower panels of each plot), integrated over the wavelength range between $2900$\,{\AA} and $5700$\,{\AA} with limb darkening coefficients derived from the 3D model (green lines) and the 1D \texttt{MARCS} model (red lines); see Table~\ref{tab:hd209458fit} for the fit parameters. The residuals of the 3D fits and the 1D fits are shown in the upper panels and center panels in each plot, the blue lines show the deviation between the 3D and 1D model light curves.}
\label{fig:hd209458whitelight}
\end{figure*}

We use the HST data of \citet{Knutsonetal:2007} for our investigation, taken as part of GO 9447 (PI Charbonneau). Results from the HD~209458b STIS dataset have also been published in \citet{Ballesteretal:2007} and \citet{Singetal:2008}. Notably, we re-reduced the HD~209458b dataset in the same manner as in \citet{Singetal:2011} for HD~189733b, producing a wavelength-integrated light curve which also has the detector position systematic trends taken into account. We find that this addition noticeably improves the quality. Transits were observed in two visits, the results of our individual fits are shown in Fig.~\ref{fig:hd209458whitelight}. The two visits together cover almost the entire light curve. Combined with the large 3D-1D effect on the limb darkening curve (cf. Fig.~\ref{fig:ldcontlineshd209458} and Fig.~\ref{fig:ld3d1dwhitelight209458}), the HD~209458 system allows for a clear distinction between the stellar model atmospheres. The improvement of the 3D model is visible in the comparison of residuals of the 3D fit and the 1D fit (black dots in upper and center panel of Fig.~\ref{fig:hd209458whitelight}), which is accompanied by a significant reduction of the $\chi^{2}$ parameter from 3D to 1D (see Table~\ref{tab:hd209458fit}). The differences between the two light curves are visualized by the blue lines in Fig.~\ref{fig:hd209458whitelight}, which closely follow the characteristic shape of the 1D fit residuals (see also Fig.~3 of \citet{Knutsonetal:2007}. The 3D-1D effect is clearly strongest in the ingress and egress phases, but also leads to a noticeable deviation around the central transit. Our result strongly indicates that the 3D model provides a more realistic description of the atmospheric temperature structure of HD~209458, which agrees well with the case of the Sun \citep{Asplundetal:2009}.

The shallower model light curve that results from overall weaker 3D limb darkening requires a slightly larger planet-to-star radius ratio to fit the observations. Averaged over the two visits in either case, the parameter increases by 0.2\,\%. Even though the overall magnitude of the 3D correction on $R_{\mathrm{Planet}}/R_{\mathrm{Star}}$ is small, the effect can still be significant for an accurate characterization of the planetary atmosphere through transmission spectroscopy, in particular if the correction is wavelength-dependent. It would therefore be interesting to investigate possible consequences for the determination of the atmospheric structure and composition of HD~209458b, but such analyses lie beyond the scope of this paper.

We note that a comparison of predicted linear and quadratic limb darkening coefficients with empirical results as in \citet{Southworth:2008} and \citet{Claret:2009} leads to similarly strong disagreement between our 3D model and the observational data, despite the clear outcome of the transit light curve fits. We suspect that inaccuracies of the linear law and also, to some degree, of the quadratic law in fitting the actual limb darkening curve and the interaction between the different free parameters of a direct fit of limb darkening are responsible for this discrepancy. \citet{Southworth:2008} indeed warns that the two coefficients of the quadratic law are correlated and therefore cannot always be uniquely determined, which inhibits a clearer comparison.

\subsection{HD~189733b}

\begin{table*}[htbp]
\caption{Parameters for the HST transit light curves of HD~189733b, shown in Fig.~\ref{fig:hd189733whitelight}.}
\begin{center}
\begin{tabular}{lllll}
\hline\hline
\multicolumn{5}{c}{HD~189733}\T\BB\\
 & \multicolumn{2}{c}{$\phantom{-}$3D} & \multicolumn{2}{c}{$\phantom{-}$1D}\\
 \hline
Parameter & $\phantom{-}$STIS & $\phantom{-}$ACS & $\phantom{-}$STIS & $\phantom{-}$ACS \T\\
\hline
$c_{1}$ & $\phantom{-}0.5598$	& $\phantom{-}0.7601$ & $\phantom{-}0.5825$ & $\phantom{-}0.6629$ \T\\
$c_{2}$ & $-0.4055$				& $-0.5528$ & $-0.7041$ & $-0.5570$ \\
$c_{3}$ & $\phantom{-}1.2498$	& $\phantom{-}0.9838$ & $\phantom{-}1.5886$ & $\phantom{-}1.0800$ \\
$c_{4}$ & $-0.4945$				& $-0.4226$ & $-0.5704$ & $-0.4474$ \\
\hline
Incl. [deg] & $\phantom{-}85.710$ & $\phantom{-}85.710$ & $\phantom{-}85.710$ & $\phantom{-}85.710$ \T\\
$\rho_{\mathrm{Star}}$\,[g\,cm$^{-3}$] & $\phantom{-}2.670$ & $\phantom{-}2.670$ & $\phantom{-}2.670$ & $\phantom{-}2.670$ \\
$a_{\mathrm{Orbit}}/R_{\mathrm{Star}}$ & $\phantom{-}8.8564$ & $\phantom{-}8.8564$ & $\phantom{-}8.8564$ & $\phantom{-}8.8564$ \\
\hline
$F_{\mathrm{Baseline}}$\,$[10^{8}\,\mathrm{counts}]$ & $\phantom{-}3.192\phantom{00}\pm0.00014$ & $\phantom{-}1.000\phantom{00}\pm0.00002$ & $\phantom{-}3.192\phantom{00}\pm0.00012$\T & $\phantom{-}1.000\phantom{00}\pm0.00003$ \\
$R_{\mathrm{Planet}}/R_{\mathrm{Star}}$ & $\phantom{-}0.15785\pm0.00023$ & $\phantom{-}0.15668\pm0.00012$ & $\phantom{-}0.15817\pm0.00020$ & $\phantom{-}0.15698\pm0.00020$ \\
\hline
DOF & $\phantom{-}89$ & $\phantom{-}107$ & $\phantom{-}89$ & $\phantom{-}107$ \T \\
$\chi^{2}$ & $\phantom{-}98$ & $\phantom{-}111$ & $\phantom{-}98$ & $\phantom{-}121$ \\
$\chi^{2}_{\nu}$ & $\phantom{-}1.10$ & $\phantom{-}1.04$ & $\phantom{-}1.10$ & $\phantom{-}1.13$ \BB \\
\hline
\end{tabular}
\tablefoot{
Fitted parameters have error bars. Parameters kept constant for different fits appear repeatedly to improve readability of the table.
}
\end{center}
\label{tab:hd189733fit}
\end{table*}

\begin{figure*}[htbp]
\centering
\includegraphics[width=8.5cm]{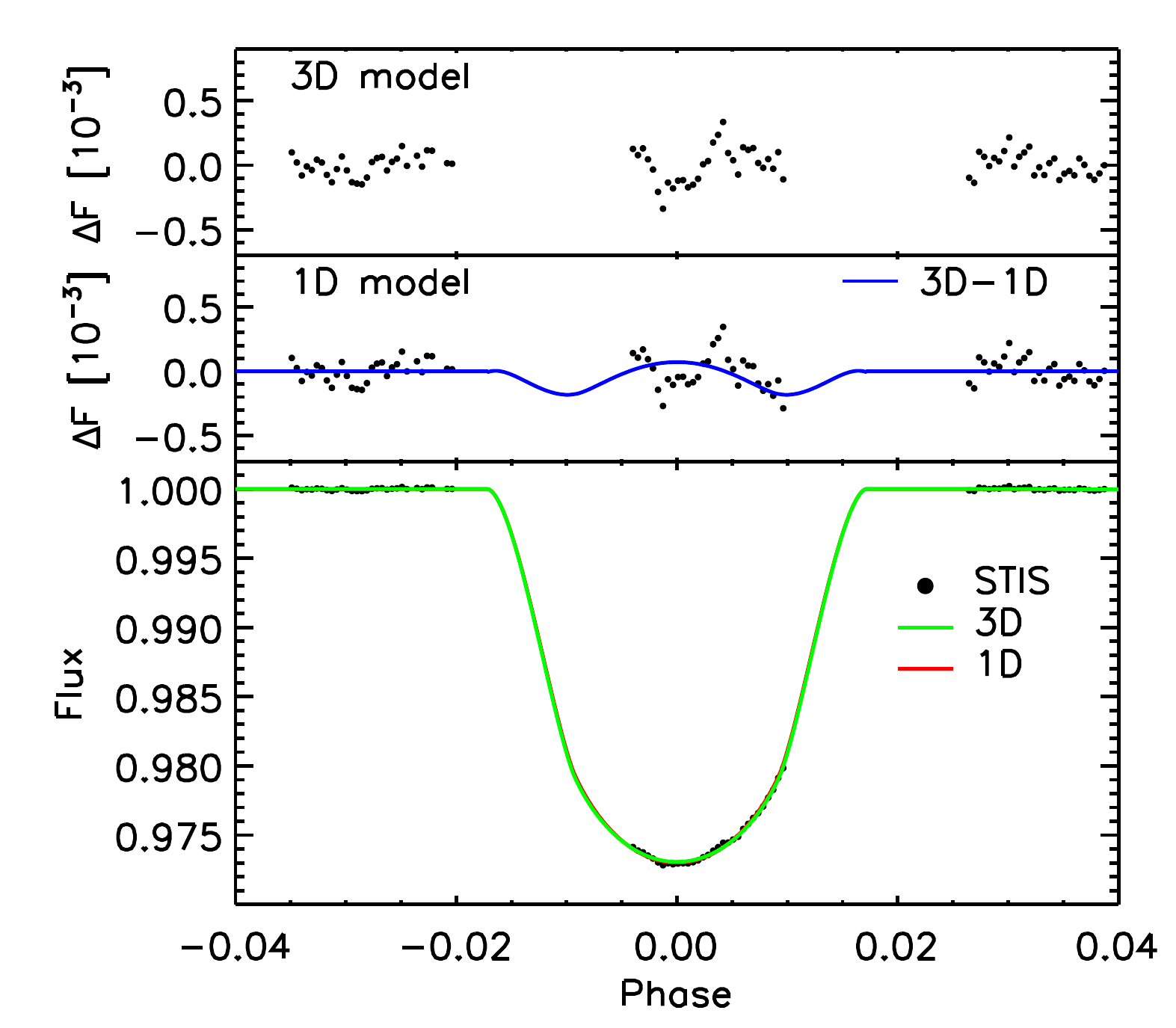}
\includegraphics[width=8.5cm]{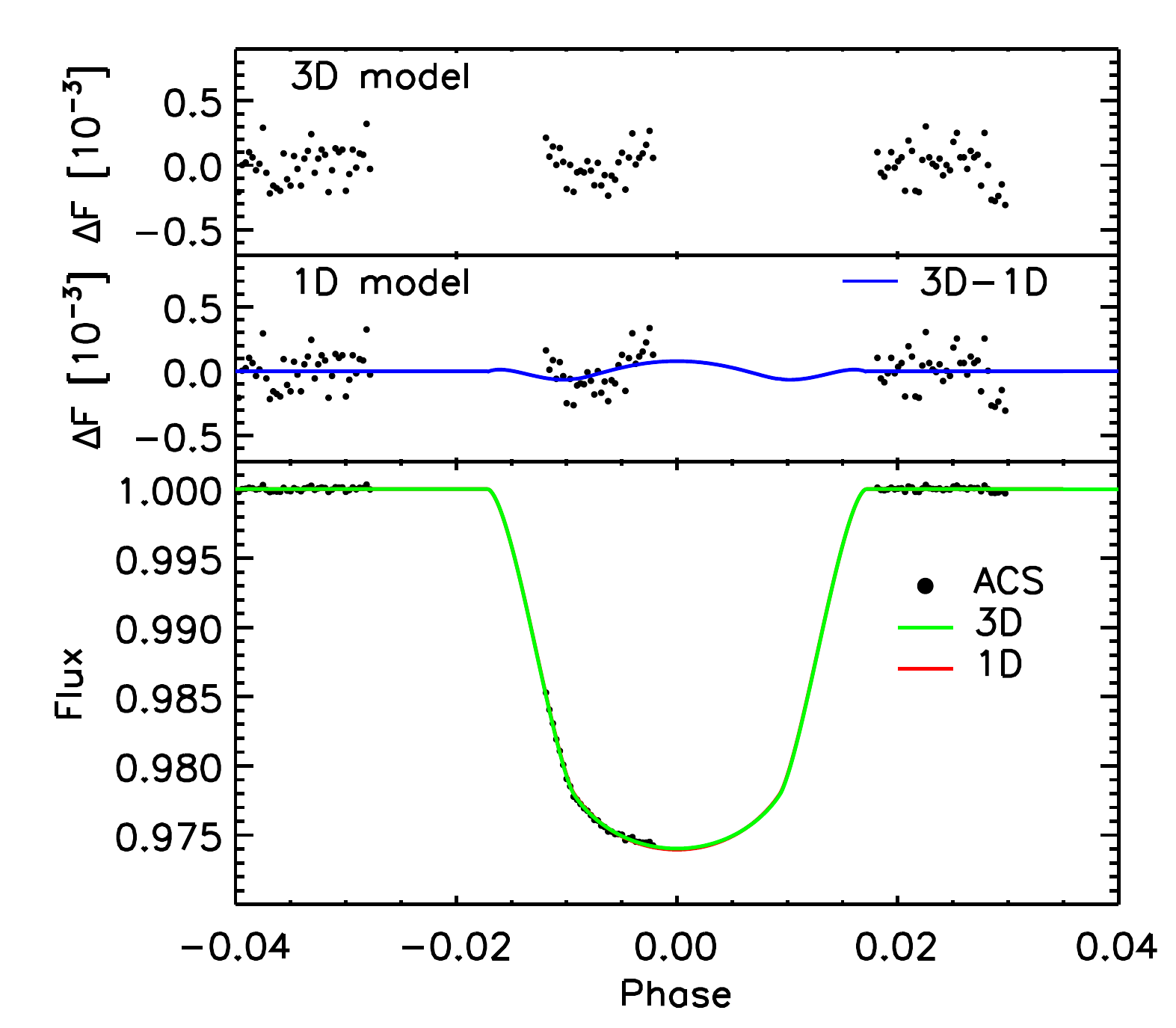}
\caption{\emph{Left:} Transit light curve fits for one HST STIS visit of HD~189733 \citep[dots in the lower panel, data from][]{Singetal:2011}, integrated over the wavelength range between $2900$\,{\AA} and $5700$\,{\AA} with limb darkening coefficients derived from the 3D model (green line) and the 1D \texttt{MARCS} model (red line); see Table~\ref{tab:hd189733fit} for the fit parameters. The residuals of the 3D fit and the 1D fit are shown in the upper panel and center panel, the blue line shows the deviation between the 3D and 1D model light curves. \emph{Right:} Transit light curve fits for one HST ACS visit of HD~189733 \cite[data from][]{Pontetal:2007}, integrated over the wavelength range between $5350$\,{\AA} and $10500$\,{\AA}.}
\label{fig:hd189733whitelight}
\end{figure*}

We use transit observations of HD~189733b from program GO 11740 (PI FP), with the data reduction method as detailed in \citet{Singetal:2011}. The HST STIS light curves are strongly distorted by star spots \citep[see Fig.~3 in][]{Singetal:2011}. While the second visit is affected in its entirety, we can use the first visit by ignoring the peak near the transit center. The observation (black dots in the lower panel on the left side of Fig.~\ref{fig:hd189733whitelight}) thus only includes a part of the egress phase which results in weaker constraints on limb darkening compared to the HD~209458 system. We also analyze the first visit of the HST ACS data of \citet{Pontetal:2007}, which is not distorted by star spots and covers parts of the ingress phase between $5350$\,{\AA} and $10500$\,{\AA} (right side of Fig.~\ref{fig:hd189733whitelight}). The model light curves based on 3D predictions are represented by green lines in the figure, red lines show the 1D \texttt{MARCS} case.

Both models reach equivalent fit quality for the STIS data (see Table~\ref{tab:hd189733fit}). The fit residuals are shown in the upper and center panels on the left side of Fig.~\ref{fig:hd189733whitelight}. The deviation between the two model light curves is represented by the blue solid line in the center panel, exhibiting again the characteristic shape of a too strong 1D limb darkening prediction. The magnitude of the effect is smaller compared to HD~209458b due to the smaller 3D-1D difference and due to stronger obstruction through spectral lines (Fig.~\ref{fig:ldcontlineshd189733} and Fig.~\ref{fig:ld3d1dwhitelight189733}). This test is not decisive enough to clearly distinguish between the temperature structures of the 3D and 1D models, as both cases deliver equivalent results with respect to the comparatively sparse observational data.

Fits of the ACS data yield a smaller $\chi^{2}$ for 3D limb darkening (see Table~\ref{tab:hd189733fit}), the residuals are shown on the right side of Fig.~\ref{fig:hd189733whitelight}. The light curve at longer wavelengths has a flatter bottom, as limb darkening is weaker due to the lower temperature-sensitivity of the Planck function. The 3D-1D difference is smaller for the same reason (blue line).

Contrary to HD~209458b, the light curves with 3D limb darkening yield slightly smaller planet-to-star radius ratios; the differences reach again 0.2\,\% for both the STIS data and ACS data.

\citet{Singetal:2011} compared light curves based on direct fits of limb darkening with a linear law and our 3D model predictions with the nonlinear law for various narrower wavelength bands within the spectral region of the STIS G430L grating. They showed that 3D limb darkening yields model light curves with similar fit quality in terms of the achieved $\chi^{2}$, even though one less free parameter is needed (see their Table~2 for details). We therefore conclude that the 3D model delivers realistic limb darkening predictions, but the 1D \texttt{MARCS} model reaches similar fit quality for the STIS data due to the low sensitivity of the test.

\section{Conclusions}\label{sec:conclusions}

We compute theoretical surface intensity distributions for the two exoplanet host stars HD~209458 and HD~189733 using 3D time-dependent hydrodynamical and 1D hydrostatic \texttt{MARCS} model atmospheres. The stellar parameters are determined using \ion{Fe}{I} and \ion{Fe}{II} abundances based on HARPS spectra, leading to slightly lower effective temperatures of the 1D models due to their different temperature stratification.

A comparison between predicted continuum intensities as a function of wavelength and radial position on the stellar disk between $0.1\le\mu\le0.9$, where $\mu$ is the cosine of the polar angle with respect to the vertical axis in the atmosphere, shows that the 3D models exhibit generally weaker limb darkening in the visual region than the 1D models, similar to an analysis of the solar atmosphere by \citet{Asplundetal:2009}. The overall shallower temperature structure of the temporal and spatial mean 3D models around the continuum optical surface in the disk center is to large parts responsible for this result: 3D models take convective motions in the atmosphere explicitly into account, rather than using a simplified prescription of convective heat flux that is inherent to 1D hydrostatic models. The 3D models exhibit slightly cooler temperatures higher up in the atmosphere compared to 1D models, which can result in a stronger 3D darkening effect close to the limb.

The presence of spectral lines partly obstructs a comparison of models if theoretical spectra are integrated over broad wavelength bands. These features form over a wide atmospheric height range and lead to a blend of limb darkening strengths that partly hides deviations in the temperature structure. It is therefore important to only compare theoretical spectra that were computed with the same set of opacities and chemical composition if the realism of stellar model atmospheres is investigated.

We analyze the results of model light curve fits to HST transit observations of the exoplanets HD~209458b and HD~189733b based on 3D and 1D limb darkening predictions. The 3D model produces a significantly better fit to the STIS light curve of HD~209458b and is capable of removing systematic residuals that appear when 1D limb darkening is used, which strongly indicates that the 3D model provides a more realistic description of the stellar temperature stratification. The improvement with 3D limb darkening is weaker for the ACS data of HD~189733b, and both models deliver equivalent fit quality for the STIS light curves. While the differences between 3D and 1D limb darkening predictions are generally smaller for this star, they are further reduced by spectral line absorption at shorter wavelengths or by the weaker limb darkening effect at longer wavelengths, respectively. A previous investigation of the same STIS data by \citet{Singetal:2011} showed that our 3D predictions rival the quality of a direct fit of limb darkening in different wavelength bands even though one free parameter less is used, which underlines the realism of our calculations.

Full intensity spectra for both stars will be made available online. However, it is important to emphasize that the opacity sampling technique used for our computations is not sufficiently accurate to represent spectral line absorption in narrow wavelength bands. Obtaining limb darkening coefficients for these cases requires a more detailed computation of stellar surface intensities with much better wavelength resolution, a detailed line list and the inclusion of Doppler shifts by the convective motions in the atmosphere. Such computations are available from the authors on request.

The weak transit signal of exoplanets with small planet-to-star radii, such as super-earth planets with solar-type host stars, will require better accuracy of the limb darkening predictions of stellar model atmospheres than what 1D models are able to provide. Already existing and future planet finding missions, such as Kepler or Plato, and the new James Webb Space Telescope (JWST), underline the importance of such efforts. A large grid of 3D hydrodynamical models is currently under construction, with surface gravities that cover the range from dwarf stars to red giant stars, effective temperatures from F-type to late K-type, and metallicities between solar and very metal-poor; see \citet{Colletetal:2011} and Magic et al. (2012, in preparation) for details. The models will soon be used for an investigation of limb darkening across a wider range of stellar parameters.

\acknowledgements{
The authors would like to thank I. Baraffe for inspiring the project, the Rechenzentrum Garching (RZG) for providing the large computational resources that were essential for this study and P. Baumann for providing co-added HARPS spectra. We would also like to thank the referee for helpful comments. W.H. acknowledges support by the European Research Council under the European Community's 7th Framework Programme (FP7/2007- 2013 Grant Agreement no. 247060). This work is based on observations with the NASA/ESA Hubble Space Telescope, obtained at the STScI operated by AURA, Inc.
}

\bibliographystyle{aa}
\bibliography{hay11}

\begin{thebibliography}{53}
\expandafter\ifx\csname natexlab\endcsname\relax\def\natexlab#1{#1}\fi

\bibitem[{{Asplund}(2005)}]{Asplund:2005}
{Asplund}, M. 2005, \araa, 43, 481

\bibitem[{{Asplund} {et~al.}(2005){Asplund}, {Grevesse}, \&
  {Sauval}}]{Asplundetal:2005}
{Asplund}, M., {Grevesse}, N., \& {Sauval}, A.~J. 2005, in Astronomical Society
  of the Pacific Conference Series, Vol. 336, Cosmic Abundances as Records of
  Stellar Evolution and Nucleosynthesis, ed. {T.~G.~Barnes III \& F.~N.~Bash},
  25

\bibitem[{{Asplund} {et~al.}(2009){Asplund}, {Grevesse}, {Sauval}, \&
  {Scott}}]{Asplundetal:2009}
{Asplund}, M., {Grevesse}, N., {Sauval}, A.~J., \& {Scott}, P. 2009, \araa, 47,
  481

\bibitem[{{Asplund} {et~al.}(2000){Asplund}, {Nordlund}, {Trampedach}, \&
  {Stein}}]{Asplundetal:2000}
{Asplund}, M., {Nordlund}, {\AA}., {Trampedach}, R., \& {Stein}, R.~F. 2000,
  \aap, 359, 743

\bibitem[{{Auvergne} {et~al.}(2009){Auvergne}, {Bodin}, {Boisnard}, {Buey},
  {Chaintreuil}, {Epstein}, {Jouret}, {Lam-Trong}, {Levacher}, {Magnan},
  {Perez}, {Plasson}, {Plesseria}, {Peter}, {Steller}, {Tiph{\`e}ne}, {Baglin},
  {Agogu{\'e}}, {Appourchaux}, {Barbet}, {Beaufort}, {Bellenger}, {Berlin},
  {Bernardi}, {Blouin}, {Boumier}, {Bonneau}, {Briet}, {Butler}, {Cautain},
  {Chiavassa}, {Costes}, {Cuvilho}, {Cunha-Parro}, {de Oliveira Fialho},
  {Decaudin}, {Defise}, {Djalal}, {Docclo}, {Drummond}, {Dupuis}, {Exil},
  {Faur{\'e}}, {Gaboriaud}, {Gamet}, {Gavalda}, {Grolleau}, {Gueguen},
  {Guivarc'h}, {Guterman}, {Hasiba}, {Huntzinger}, {Hustaix}, {Imbert},
  {Jeanville}, {Johlander}, {Jorda}, {Journoud}, {Karioty}, {Kerjean},
  {Lafond}, {Lapeyrere}, {Landiech}, {Larqu{\'e}}, {Laudet}, {Le Merrer},
  {Leporati}, {Leruyet}, {Levieuge}, {Llebaria}, {Martin}, {Mazy}, {Mesnager},
  {Michel}, {Moalic}, {Monjoin}, {Naudet}, {Neukirchner}, {Nguyen-Kim},
  {Ollivier}, {Orcesi}, {Ottacher}, {Oulali}, {Parisot}, {Perruchot},
  {Piacentino}, {Pinheiro da Silva}, {Platzer}, {Pontet}, {Pradines},
  {Quentin}, {Rohbeck}, {Rolland}, {Rollenhagen}, {Romagnan}, {Russ}, {Samadi},
  {Schmidt}, {Schwartz}, {Sebbag}, {Smit}, {Sunter}, {Tello}, {Toulouse},
  {Ulmer}, {Vandermarcq}, {Vergnault}, {Wallner}, {Waultier}, \&
  {Zanatta}}]{Auvergneetal:2009}
{Auvergne}, M., {Bodin}, P., {Boisnard}, L., {et~al.} 2009, \aap, 506, 411

\bibitem[{{Ballester} {et~al.}(2007){Ballester}, {Sing}, \&
  {Herbert}}]{Ballesteretal:2007}
{Ballester}, G.~E., {Sing}, D.~K., \& {Herbert}, F. 2007, \nat, 445, 511

\bibitem[{{Blackwell} {et~al.}(1995){Blackwell}, {Lynas-Gray}, \&
  {Smith}}]{Blackwelletal:1995}
{Blackwell}, D.~E., {Lynas-Gray}, A.~E., \& {Smith}, G. 1995, \aap, 296, 217

\bibitem[{{Bouchy} {et~al.}(2005){Bouchy}, {Udry}, {Mayor}, {Moutou}, {Pont},
  {Iribarne}, {da Silva}, {Ilovaisky}, {Queloz}, {Santos}, {S{\'e}gransan}, \&
  {Zucker}}]{Bouchyetal:2005}
{Bouchy}, F., {Udry}, S., {Mayor}, M., {et~al.} 2005, \aap, 444, L15

\bibitem[{{Brown} {et~al.}(2001){Brown}, {Charbonneau}, {Gilliland}, {Noyes},
  \& {Burrows}}]{Brownetal:2001}
{Brown}, T.~M., {Charbonneau}, D., {Gilliland}, R.~L., {Noyes}, R.~W., \&
  {Burrows}, A. 2001, \apj, 552, 699

\bibitem[{{Casagrande} {et~al.}(2011){Casagrande}, {Sch{\"o}nrich}, {Asplund},
  {Cassisi}, {Ram{\'{\i}}rez}, {Mel{\'e}ndez}, {Bensby}, \&
  {Feltzing}}]{Casagrandeetal:2011}
{Casagrande}, L., {Sch{\"o}nrich}, R., {Asplund}, M., {et~al.} 2011, \aap, 530,
  A138

\bibitem[{{Chiavassa} {et~al.}(2009){Chiavassa}, {Plez}, {Josselin}, \&
  {Freytag}}]{Chiavassaetal:2009}
{Chiavassa}, A., {Plez}, B., {Josselin}, E., \& {Freytag}, B. 2009, \aap, 506,
  1351

\bibitem[{{Claret}(2000)}]{Claret:2000}
{Claret}, A. 2000, \aap, 363, 1081

\bibitem[{{Claret}(2009)}]{Claret:2009}
{Claret}, A. 2009, \aap, 506, 1335

\bibitem[{{Claret} \& {Hauschildt}(2003)}]{Claretetal:2003}
{Claret}, A. \& {Hauschildt}, P.~H. 2003, \aap, 412, 241

\bibitem[{{Collet} {et~al.}(2011){Collet}, {Magic}, \&
  {Asplund}}]{Colletetal:2011}
{Collet}, R., {Magic}, Z., \& {Asplund}, M. 2011, Journal of Physics Conference
  Series, 328, 012003

\bibitem[{{Gonzalez} {et~al.}(2001){Gonzalez}, {Laws}, {Tyagi}, \&
  {Reddy}}]{Gonzalezetal:2001}
{Gonzalez}, G., {Laws}, C., {Tyagi}, S., \& {Reddy}, B.~E. 2001, \aj, 121, 432

\bibitem[{{Gustafsson} {et~al.}(1975){Gustafsson}, {Bell}, {Eriksson}, \&
  {Nordlund}}]{Gustafssonetal:1975}
{Gustafsson}, B., {Bell}, R.~A., {Eriksson}, K., \& {Nordlund}, A. 1975, \aap,
  42, 407

\bibitem[{{Gustafsson} {et~al.}(2008){Gustafsson}, {Edvardsson}, {Eriksson},
  {J{\o}rgensen}, {Nordlund}, \& {Plez}}]{Gustafssonetal:2008}
{Gustafsson}, B., {Edvardsson}, B., {Eriksson}, K., {et~al.} 2008, \aap, 486,
  951

\bibitem[{{Hauschildt} {et~al.}(1999){Hauschildt}, {Allard}, \&
  {Baron}}]{Hauschildtetal:1999}
{Hauschildt}, P.~H., {Allard}, F., \& {Baron}, E. 1999, \apj, 512, 377

\bibitem[{{Hayek} {et~al.}(2010){Hayek}, {Asplund}, {Carlsson}, {Trampedach},
  {Collet}, {Gudiksen}, {Hansteen}, \& {Leenaarts}}]{Hayeketal:2010}
{Hayek}, W., {Asplund}, M., {Carlsson}, M., {et~al.} 2010, \aap, 517, A49

\bibitem[{{Hayek} {et~al.}(2011){Hayek}, {Asplund}, {Collet}, \&
  {Nordlund}}]{Hayeketal:2011}
{Hayek}, W., {Asplund}, M., {Collet}, R., \& {Nordlund}, {\AA}. 2011, \aap,
  529, A158

\bibitem[{{Heiter} \& {Luck}(2003)}]{Heiteretal:2003}
{Heiter}, U. \& {Luck}, R.~E. 2003, \aj, 126, 2015

\bibitem[{{Heyrovsk{\'y}}(2007)}]{Heyrovsky:2007}
{Heyrovsk{\'y}}, D. 2007, \apj, 656, 483

\bibitem[{{Holweger} {et~al.}(1995){Holweger}, {Kock}, \&
  {Bard}}]{Holwegeretal:1995}
{Holweger}, H., {Kock}, M., \& {Bard}, A. 1995, \aap, 296, 233

\bibitem[{{Knutson} {et~al.}(2007){Knutson}, {Charbonneau}, {Noyes}, {Brown},
  \& {Gilliland}}]{Knutsonetal:2007}
{Knutson}, H.~A., {Charbonneau}, D., {Noyes}, R.~W., {Brown}, T.~M., \&
  {Gilliland}, R.~L. 2007, \apj, 655, 564

\bibitem[{{Kurucz}(1996)}]{Kurucz:1996}
{Kurucz}, R.~L. 1996, in IAU Symposium, Vol. 176, Stellar Surface Structure,
  ed. {K.~G.~Strassmeier \& J.~L.~Linsky}, 523

\bibitem[{{Mandel} \& {Agol}(2002)}]{Mandeletal:2002}
{Mandel}, K. \& {Agol}, E. 2002, \apjl, 580, L171

\bibitem[{{Mashonkina} \& {Gehren}(2001)}]{Mashonkinaetal:2001}
{Mashonkina}, L. \& {Gehren}, T. 2001, \aap, 376, 232

\bibitem[{{Mazeh} {et~al.}(2000){Mazeh}, {Naef}, {Torres}, {Latham}, {Mayor},
  {Beuzit}, {Brown}, {Buchhave}, {Burnet}, {Carney}, {Charbonneau}, {Drukier},
  {Laird}, {Pepe}, {Perrier}, {Queloz}, {Santos}, {Sivan}, {Udry}, \&
  {Zucker}}]{Mazehetal:2000}
{Mazeh}, T., {Naef}, D., {Torres}, G., {et~al.} 2000, \apjl, 532, L55

\bibitem[{{Mel{\'e}ndez} \& {Barbuy}(2009)}]{Melendezetal:2009}
{Mel{\'e}ndez}, J. \& {Barbuy}, B. 2009, \aap, 497, 611

\bibitem[{{Mihalas}(1978)}]{Mihalas:1978}
{Mihalas}, D. 1978, {Stellar atmospheres, 2nd edition}, ed. J.~Hevelius (San
  Francisco: W.~H.~Freeman and Co.)

\bibitem[{{Neckel} \& {Labs}(1994)}]{Neckeletal:1994}
{Neckel}, H. \& {Labs}, D. 1994, \solphys, 153, 91

\bibitem[{{Nordlund}(1982)}]{Nordlund:1982}
{Nordlund}, A. 1982, \aap, 107, 1

\bibitem[{Nordlund \& Galsgaard(1995)}]{Nordlundetal:1995}
Nordlund, {\AA}. \& Galsgaard, K. 1995, A 3D {MHD} Code for Parallel Computers,
  Tech. rep., Astronomical Observatory, Copenhagen University

\bibitem[{{Nordlund} {et~al.}(2009){Nordlund}, {Stein}, \&
  {Asplund}}]{Nordlundetal:2009}
{Nordlund}, {\AA}., {Stein}, R.~F., \& {Asplund}, M. 2009, Living Reviews in
  Solar Physics, 6, 2

\bibitem[{{Pereira} {et~al.}(2009{\natexlab{a}}){Pereira}, {Asplund}, \&
  {Kiselman}}]{Pereiraetal:2009b}
{Pereira}, T.~M.~D., {Asplund}, M., \& {Kiselman}, D. 2009{\natexlab{a}}, \aap,
  508, 1403

\bibitem[{{Pereira} {et~al.}(2009{\natexlab{b}}){Pereira}, {Kiselman}, \&
  {Asplund}}]{Pereiraetal:2009a}
{Pereira}, T.~M.~D., {Kiselman}, D., \& {Asplund}, M. 2009{\natexlab{b}}, \aap,
  507, 417

\bibitem[{{Plez}(2008)}]{Plez:2008}
{Plez}, B. 2008, Physica Scripta Volume T, 133, 014003

\bibitem[{{Pont} {et~al.}(2007){Pont}, {Gilliland}, {Moutou}, {Charbonneau},
  {Bouchy}, {Brown}, {Mayor}, {Queloz}, {Santos}, \& {Udry}}]{Pontetal:2007}
{Pont}, F., {Gilliland}, R.~L., {Moutou}, C., {et~al.} 2007, \aap, 476, 1347

\bibitem[{{Pont} {et~al.}(2008){Pont}, {Knutson}, {Gilliland}, {Moutou}, \&
  {Charbonneau}}]{Pontetal:2008}
{Pont}, F., {Knutson}, H., {Gilliland}, R.~L., {Moutou}, C., \& {Charbonneau},
  D. 2008, \mnras, 385, 109

\bibitem[{{Ram{\'{\i}}rez} \& {Mel{\'e}ndez}(2004)}]{Ramirezetal:2004}
{Ram{\'{\i}}rez}, I. \& {Mel{\'e}ndez}, J. 2004, \apj, 609, 417

\bibitem[{{Sadakane} {et~al.}(2002){Sadakane}, {Ohkubo}, {Takeda}, {Sato},
  {Kambe}, \& {Aoki}}]{Sadakaneetal:2002}
{Sadakane}, K., {Ohkubo}, M., {Takeda}, Y., {et~al.} 2002, \pasj, 54, 911

\bibitem[{{Santos} {et~al.}(2004){Santos}, {Israelian}, \&
  {Mayor}}]{Santosetal:2004}
{Santos}, N.~C., {Israelian}, G., \& {Mayor}, M. 2004, \aap, 415, 1153

\bibitem[{{Sing}(2010)}]{Sing:2010}
{Sing}, D.~K. 2010, \aap, 510, A21

\bibitem[{{Sing} {et~al.}(2011){Sing}, {Pont}, {Aigrain}, {Charbonneau},
  {D{\'e}sert}, {Gibson}, {Gilliland}, {Hayek}, {Henry}, {Knutson}, {Lecavelier
  Des Etangs}, {Mazeh}, \& {Shporer}}]{Singetal:2011}
{Sing}, D.~K., {Pont}, F., {Aigrain}, S., {et~al.} 2011, \mnras, 1159

\bibitem[{{Sing} {et~al.}(2008){Sing}, {Vidal-Madjar}, {D{\'e}sert},
  {Lecavelier des Etangs}, \& {Ballester}}]{Singetal:2008}
{Sing}, D.~K., {Vidal-Madjar}, A., {D{\'e}sert}, J., {Lecavelier des Etangs},
  A., \& {Ballester}, G. 2008, \apj, 686, 658

\bibitem[{{Skartlien}(2000)}]{Skartlien:2000}
{Skartlien}, R. 2000, \apj, 536, 465

\bibitem[{{Sousa} {et~al.}(2008){Sousa}, {Santos}, {Mayor}, {Udry},
  {Casagrande}, {Israelian}, {Pepe}, {Queloz}, \& {Monteiro}}]{Sousaetal:2008}
{Sousa}, S.~G., {Santos}, N.~C., {Mayor}, M., {et~al.} 2008, \aap, 487, 373

\bibitem[{{Southworth}(2008)}]{Southworth:2008}
{Southworth}, J. 2008, \mnras, 386, 1644

\bibitem[{{Southworth}(2009)}]{Southworth:2009}
{Southworth}, J. 2009, \mnras, 394, 272

\bibitem[{{Southworth}(2010)}]{Southworth:2010}
{Southworth}, J. 2010, \mnras, 408, 1689

\bibitem[{{Valenti} \& {Fischer}(2005)}]{Valentietal:2005}
{Valenti}, J.~A. \& {Fischer}, D.~A. 2005, \apjs, 159, 141

\bibitem[{{van Belle} \& {von Braun}(2009)}]{vanBelleetal:2009}
{van Belle}, G.~T. \& {von Braun}, K. 2009, \apj, 694, 1085

\end{thebibliography}

\appendix

\section{Limb darkening coefficients for standard broadband filters and different instruments}\label{sec:ldccompilation}

\begin{table*}[htdp]
\caption{Theoretical limb darkening coefficients for a range of standard broadband filters and instruments.}
\begin{center}
\begin{tabular}{l|r|rrrr|rrrr}
\hline
\hline
\multicolumn{1}{c}{} & \multicolumn{1}{c}{} & \multicolumn{4}{c}{HD~209458} & \multicolumn{4}{c}{HD~189733} \\
\multicolumn{1}{c}{Band} & \multicolumn{1}{c}{$\lambda_{\mathrm{eff}}$} & \multicolumn{1}{c}{$c_{1}$} & \multicolumn{1}{c}{$c_{2}$} & \multicolumn{1}{c}{$c_{3}$} & \multicolumn{1}{c}{$c_{4}$} & \multicolumn{1}{c}{$c_{1}$} & \multicolumn{1}{c}{$c_{2}$} & \multicolumn{1}{c}{$c_{3}$} & \multicolumn{1}{c}{$c_{4}$} \\
\hline
    Johnson U & $ 3614$\,{\AA} & $0.2714$ & $ 0.6757$ & $-0.0135$ & $-0.0423$ & $0.6187$ & $-0.9692$ & $1.7593$ & $-0.4501$ \\
    Johnson B & $ 4414$\,{\AA} & $0.4089$ & $ 0.4379$ & $ 0.0826$ & $-0.0591$ & $0.5269$ & $-0.4716$ & $1.4046$ & $-0.5285$ \\
    Johnson V & $ 5514$\,{\AA} & $0.5907$ & $ 0.1358$ & $ 0.1310$ & $-0.0603$ & $0.6198$ & $-0.3657$ & $1.0856$ & $-0.4616$ \\
    Johnson I & $ 8867$\,{\AA} & $0.7463$ & $-0.3785$ & $ 0.4081$ & $-0.1494$ & $0.7857$ & $-0.6139$ & $0.9500$ & $-0.4035$ \\
\hline
Str\"omgren u & $ 3467$\,{\AA} & $0.2489$ & $ 0.7509$ & $-0.0881$ & $-0.0130$ & $0.5627$ & $-0.9727$ & $1.8312$ & $-0.4517$ \\
Str\"omgren v & $ 4110$\,{\AA} & $0.3302$ & $ 0.5697$ & $ 0.0501$ & $-0.0583$ & $0.5298$ & $-0.7119$ & $1.7180$ & $-0.5853$ \\
Str\"omgren b & $ 4673$\,{\AA} & $0.4577$ & $ 0.4220$ & $ 0.0033$ & $-0.0220$ & $0.4970$ & $-0.3015$ & $1.2544$ & $-0.5223$ \\
Str\"omgren y & $ 5479$\,{\AA} & $0.5958$ & $ 0.1268$ & $ 0.1376$ & $-0.0624$ & $0.6171$ & $-0.3405$ & $1.0592$ & $-0.4560$ \\
\hline
      SDSS u' & $ 3572$\,{\AA} & $0.2616$ & $ 0.7449$ & $-0.1043$ & $-0.0105$ & $0.6085$ & $-0.9915$ & $1.8023$ & $-0.4572$ \\
      SDSS g' & $ 4751$\,{\AA} & $0.4692$ & $ 0.3707$ & $ 0.0510$ & $-0.0408$ & $0.5316$ & $-0.3781$ & $1.2733$ & $-0.5102$ \\
      SDSS r' & $ 6204$\,{\AA} & $0.6716$ & $-0.0532$ & $ 0.2233$ & $-0.0901$ & $0.6967$ & $-0.4221$ & $1.0160$ & $-0.4450$ \\
      SDSS i' & $ 7519$\,{\AA} & $0.7341$ & $-0.2850$ & $ 0.3645$ & $-0.1369$ & $0.7725$ & $-0.5732$ & $1.0026$ & $-0.4293$ \\
      SDSS z' & $ 8992$\,{\AA} & $0.7505$ & $-0.4067$ & $ 0.4254$ & $-0.1552$ & $0.7930$ & $-0.6326$ & $0.9467$ & $-0.4002$ \\
\hline
    Bessell J & $1.241$\,$\mu$m & $0.7277$ & $-0.3488$ & $ 0.2343$ & $-0.0643$ & $0.7422$ & $-0.4298$ & $0.5875$ & $-0.2645$ \\
    Bessell H & $1.651$\,$\mu$m & $1.0576$ & $-1.1056$ & $ 0.8110$ & $-0.2321$ & $0.9618$ & $-0.6133$ & $0.4033$ & $-0.1357$ \\
    Bessell K & $2.218$\,$\mu$m & $0.8614$ & $-0.8792$ & $ 0.6492$ & $-0.1899$ & $0.9065$ & $-0.7574$ & $0.5863$ & $-0.2043$ \\
\hline
           MK J & $ 1.250 $\,$\mu$m & $ 0.7263 $ & $ -0.3442 $ & $  0.2243 $ & $ -0.0600  $ & $ 0.7355 $ & $ -0.4058 $ & $  0.5573 $ & $ -0.2537  $ \\
           MK H & $ 1.634 $\,$\mu$m & $ 1.0481 $ & $ -1.0749 $ & $  0.7813 $ & $ -0.2218  $ & $ 0.9488 $ & $ -0.5850 $ & $  0.3856 $ & $ -0.1318  $ \\
          MK K' & $ 2.118 $\,$\mu$m & $ 0.8959 $ & $ -0.9385 $ & $  0.7010 $ & $ -0.2062  $ &  $ 0.9193 $ & $ -0.7365 $ & $  0.5446 $ & $ -0.1860  $ \\
          MK Ks & $ 2.149 $\,$\mu$m & $ 0.8805 $ & $ -0.9149 $ & $  0.6833 $ & $ -0.2014  $ & $ 0.9133 $ & $ -0.7368 $ & $  0.5485 $ & $ -0.1878  $ \\
           MK K & $ 2.205 $\,$\mu$m & $ 0.8617 $ & $ -0.8822 $ & $  0.6530 $ & $ -0.1914  $ & $ 0.9061 $ & $ -0.7508 $ & $  0.5752 $ & $ -0.1995  $ \\
          MK L' & $ 3.785 $\,$\mu$m & $ 0.5370 $ & $ -0.5436 $ & $  0.4407 $ & $ -0.1423  $ & $ 0.6874 $ & $ -0.7468 $ & $  0.6881 $ & $ -0.2531  $ \\
          MK M' & $ 4.687 $\,$\mu$m & $ 0.4241 $ & $ -0.3318 $ & $  0.2333 $ & $ -0.0696  $ & $ 0.6878 $ & $ -0.9198 $ & $  0.8999 $ & $ -0.3344  $ \\
\hline
STIS G430L\tablefootmark{1} & $ 4757$\,{\AA} & $0.4762$ & $ 0.3436$ & $ 0.0730$ & $-0.0492$ & $0.5601$ & $-0.4054$ & $1.2492$ & $-0.4943$ \\
ACS HRC G800L\tablefootmark{1} & $ 7945$\,{\AA} & $ 0.7112$ & $-0.2198$ & $ 0.2926$ & $-0.1070$ & $ 0.7601$ & $-0.5527$ & $ 0.9838$ & $-0.4226$ \\
\hline
         Kepler & $ 6641$\,{\AA} & $0.6434$ & $-0.0242$ & $ 0.2217$ & $-0.0909$ & $0.6890$ & $-0.4600$ & $1.0493$ & $-0.4488$ \\
\hline
          Corot & $ 6958$\,{\AA} & $0.6435$ & $-0.0410$ & $ 0.2360$ & $-0.0961$ & $0.6987$ & $-0.4833$ & $1.0479$ & $-0.4451$ \\
\hline
 Spitzer IRAC 1 & $3.576$\,$\mu$m & $0.5564$ & $-0.5462$ & $ 0.4315$ & $-0.1368$ & $0.7062$ & $-0.7408$ & $0.6723$ & $-0.2464$ \\
 Spitzer IRAC 2 & $4.529$\,$\mu$m & $0.4614$ & $-0.4277$ & $ 0.3362$ & $-0.1074$ & $0.6830$ & $-0.8708$ & $0.8418$ & $-0.3121$ \\
 Spitzer IRAC 3 & $5.788$\,$\mu$m & $0.4531$ & $-0.5119$ & $ 0.4335$ & $-0.1431$ & $0.6290$ & $-0.8773$ & $0.8715$ & $-0.3256$ \\
 Spitzer IRAC 4 & $8.045$\,$\mu$m & $0.4354$ & $-0.6067$ & $ 0.5421$ & $-0.1816$ & $0.6189$ & $-0.9742$ & $0.9804$ & $-0.3664$ \\
\hline
\end{tabular}
\tablefoot{All computations are based on the 3D models of HD~209458 and HD~189733. \tablefoottext{1}{Theoretical intensity spectra were integrated over the entire accessible wavelength range of the grating, leading to small deviations from the coefficients in Table~\ref{tab:hd209458fit} and Table~\ref{tab:hd189733fit}.}
}
\end{center}
\label{tab:ldccompilation}
\end{table*}

Limb darkening coefficients for the \citet{Claret:2000} law were computed for a range of standard broadband filters and instruments, based on the 3D models of HD~209458 and HD~189733. The coefficients are derived from least-squares fits to the wavelength-integrated limb darkening law,
\begin{equation}
\frac{I(\mu)}{I(1)}=\frac{\int\lambda S_{\lambda}I_{\lambda}(\mu)d\lambda}{\int\lambda S_{\lambda}I_{\lambda}(1)d\lambda},
\end{equation}
where $S_{\lambda}$ is the filter or instrument response curve, respectively, and the factor $\lambda$ converts the incoming intensity $I_{\lambda}$ into a photon flux; see also Sect.~\ref{sec:specld}. The effective wavelength of each filter was calculated using the integral
\begin{equation}
\lambda_{\mathrm{eff}}=\frac{\int\lambda^{2}S_{\lambda}d\lambda}{\int\lambda S_{\lambda}d\lambda}.
\end{equation}
Table~\ref{tab:ldccompilation} gives a summary of the results. Standard filter curves were taken from the \texttt{SYNPHOT} package (\texttt{SYNPHOT} is a product of the Space Telescope Science Institute, which is operated by AURA for NASA). The New Mauna Kea (MK) filter set was downloaded from irtfweb.ifa.hawaii.edu/$\sim$nsfcam/filters.html. The STIS and ACS sensitivity curves were downloaded from the STScI calibration database. The Kepler sensitivity curve was downloaded from the Kepler website (keplergo.arc.nasa.gov/kepler\_response\_hires1.txt). The Corot sensitivity curve was extracted from \citet{Auvergneetal:2009}. The Spitzer sensitivity curves were downloaded from irsa.ipac.caltech.edu/data/SPITZER/docs/irac/calibrationfiles/spectralresponse/.

\section{Dependence of limb darkening on the atmospheric temperature gradient}\label{sec:ldtttau}

The strength of limb darkening is closely related to the vertical atmospheric temperature gradient near the optical surface and depends on wavelength, which will be established in the following using basic stellar atmosphere theory \citep[see, e.g.,][]{Mihalas:1978}. In the Eddington-Barbier approximation, the depth-dependence of the Planck function is simplified by the linear expression
\begin{equation}
B_{\lambda}(\tau_{\lambda})\approx a_{\lambda}+b_{\lambda}\tau_{\lambda},
\label{equ:eddbarbsrc}
\end{equation}
where $\tau_{\lambda}$ is the vertical monochromatic optical depth with $\tau_{\lambda}=0$ outside the star, and $\tau_{\lambda}=1$ marks the optical surface at wavelength $\lambda$. $a_{\lambda}$ and $b_{\lambda}$ are constants, which can be expressed as
\begin{equation}
b_{\lambda}=\frac{dB_{\lambda}}{d\tau_{\lambda}},
\label{eqn:blambda}
\end{equation}
and
\begin{equation}
a_{\lambda}=B_{\lambda}(\tau_{\lambda}=1)-b_{\lambda}=B_{\lambda}(\tau_{\lambda}=1)-\frac{dB_{\lambda}}{d\tau_{\lambda}},
\label{eqn:alambda}
\end{equation}
where $B_{\lambda}(\tau_{\lambda}=1)$ is the Planck function at the optical surface in the stellar disk center. In plane-parallel geometry, the source function (Eq.~(\ref{equ:eddbarbsrc})) produces the surface radiation field
\begin{equation}
I_{\lambda}(\mu)=a_{\lambda}+b_{\lambda}\mu=B_{\lambda}(\tau_{\lambda}=\mu),
\label{eqn:surfint}
\end{equation}
which has linear anisotropy and consequently leads to a linear limb darkening relation. The second equality means that the surface intensity observed at angle $\mu$ is emitted from vertical optical depth $\tau_{\lambda}=\mu$ in the atmosphere. Inserting the constants $a_{\lambda}$ and $b_{\lambda}$, one obtains
\begin{equation}
I_{\lambda}(\mu)=B_{\lambda}(\tau_{\lambda}=1)-\frac{dB_{\lambda}}{d\tau_{\lambda}}\left[1-\mu\right],
\end{equation}
which is equivalent to a limb darkening law
\begin{equation}
\frac{I_{\lambda}(\mu)}{I_{\lambda}(1)}=1-\frac{1}{B_{\lambda}(\tau_{\lambda}=1)}\frac{dB_{\lambda}}{d\tau_{\lambda}}\left[1-\mu\right]\equiv1-u_{\lambda}\left[1-\mu\right].
\end{equation}
Using the chain rule, one can express the strength $u_{\lambda}$ of of the limb darkening effect through
\begin{equation}
u_{\lambda}=\frac{1}{B_{\lambda}(\tau_{\lambda}=1)}\frac{dB_{\lambda}}{d\tau_{\lambda}}=\frac{\log e}{B_{\lambda}(T(\tau_{\lambda}=1))}\left.\frac{dB_{\lambda}}{dT}\right|_{\tau_{\lambda}=1}\left.\frac{dT}{d\log\tau_{\lambda}}\right|_{\tau_{\lambda}=1},
\label{eqn:ldccoeff}
\end{equation}
where $dT/d\log\tau_{\lambda}$ is the vertical atmospheric temperature gradient at the monochromatic optical surface $\tau_{\lambda}=1$; the logarithmic optical depth is commonly used as a depth variable and facilitates comparison with Fig.~\ref{fig:3Dmodel}. A steeper gradient at the optical surface will therefore produce a stronger limb darkening effect. Likewise, if the opacity (and thus the atmospheric height of the optical surface) varies only weakly with wavelength, the increasing relative temperature sensitivity of the Planck function, expressed through the factor $(dB_{\lambda}/dT)/B_{\lambda}$, increases the limb darkening effect monotonously towards shorter wavelengths.

The dependence of limb darkening for a given atmospheric stratification on monochromatic opacity, which can vary strongly between a spectral line core and the continuum, can be understood with similar arguments. Assuming a second opacity at nearly the same wavelength that scales linearly with respect to the first one, independent of depth,
\begin{equation}
\chi'_{\lambda}=k\chi_{\lambda},
\end{equation}
leads to the relation
\begin{equation}
\tau'_{\lambda}=k\tau_{\lambda}
\label{eqn:tautaudash}
\end{equation}
between the two optical depths. Using Eq.~(\ref{eqn:ldccoeff}) and Eq.~(\ref{eqn:tautaudash}), the limb darkening strength $u'$ is given by
\begin{equation}
u'_{\lambda}=\frac{1}{B_{\lambda}(\tau'_{\lambda}=1)}\frac{dB_{\lambda}}{d\tau'_{\lambda}}=\frac{1}{B_{\lambda}(\tau_{\lambda}=1/k)}\frac{dB_{\lambda}}{d\tau_{\lambda}}\frac{1}{k}.
\label{eqn:ldcdoeffdash}
\end{equation}
For $k>1$, the optical surface $\tau'_{\lambda}=1$ moves to larger atmospheric height at $\tau_{\lambda}=1/k$ on the original scale where thermal emission $B_{\lambda}$ is smaller if gas temperature decreases monotonously outward. At the same time, the source function gradient $dB_{\lambda}/d\tau'_{\lambda}$ decreases by a factor of $k$ with respect to the original gradient $dB_{\lambda}/d\tau_{\lambda}$. Using Eq.~(\ref{equ:eddbarbsrc}) through Eq.~(\ref{eqn:alambda}) and Eq.~(\ref{eqn:ldccoeff}), the source function is expressed in terms of $u_{\lambda}$:
\begin{equation}
B_{\lambda}(\tau_{\lambda}=1/k)=B_{\lambda}(\tau_{\lambda}=1)-\frac{dB_{\lambda}}{d\tau_{\lambda}}\left[1-\frac{1}{k}\right]=B_{\lambda}(\tau_{\lambda}=1)\left\{1-u_{\lambda}\left[1-\frac{1}{k}\right]\right\}.
\label{eqn:thermsourcedash}
\end{equation}
Inserting this result into Eq.~(\ref{eqn:ldcdoeffdash}), one obtains the simple relation
\begin{equation}
u'_{\lambda}=\frac{u_{\lambda}}{1-u_{\lambda}\left[1-\frac{1}{k}\right]}\frac{1}{k}.
\end{equation}
It is easy to verify that $u'_{\lambda}\le u_{\lambda}$ for $0\le u_{\lambda}\le1$ and $k>1$. $u'_{\lambda}=u_{\lambda}$ only holds in the limits of an isothermal atmosphere ($u_{\lambda}=0$) or with maximal limb darkening and zero emission at the limb ($u_{\lambda}=1$). Between these extremes, limb darkening is always weaker in line cores than in the continuum, if temperature decreases monotonously with increasing atmospheric height.

\end{document}